\documentstyle{article}
\begin{document}

\newcommand{\leaveout}[1]{}   
\newcommand{\leavein}[1]{#1}
\newcommand{\un}{u^{(n)}}     
\newcommand{\vn}{v^{(n)}}
\newcommand{\An}{A^{(n)}}
\newcommand{\Yn}{Y^{(n)}}
\newcommand{\Sigman}{\Sigma^{(n)}}
\newcommand{\Rn}{R^{(n)}}
\newcommand{\rhon}{\rho^{(n)}}
\newcommand{\psin}{\psi^{(n)}}
\newcommand{\phin}{\phi^{(n)}}
\newcommand{\mun}{\mu^{(n)}}
\newcommand{\betan}{\beta^{(n)}}
\newcommand{\lambdan}{\lambda^{(n)}}
\newcommand{\unmean}{\langle u^{(n)} \rangle}
\newcommand{\vnmean}{\langle v^{(n)} \rangle}
\newcommand{\unxmean}{\langle u^{(n)}(x) \rangle}
\newcommand{\vnxmean}{\langle v^{(n)}(x) \rangle}
\newcommand{\ungrad}{\langle u^{(n)}_{x} \rangle}
\newcommand{\vngrad}{\langle v^{(n)}_{x} \rangle}
\newcommand{\unxgrad}{\langle u^{(n)}_{x}(x) \rangle}
\newcommand{\vnxgrad}{\langle v^{(n)}_{x}(x) \rangle}

\title{A mean--field statistical theory for the nonlinear
  Schr\"odinger equation}
\author{Richard Jordan \footnote{Corresponding author:
phone: (505) 667-8715; fax: (505) 665-2659; 
e-mail: rjordan@cnls.lanl.gov} \\
 T-7 and CNLS\\ MS B258\\Los Alamos National
Laboratory\\ Los Alamos, NM 87545
\and Bruce Turkington \\ Department of Mathematics and Statistics \\
University of Massachusetts\\ Amherst, MA 01003
\and Craig L. Zirbel \\ Department of Mathematics and Statistics \\
Bowling Green State University\\ Bowling Green, OH 43403}

\date{April 5, 1999}

\maketitle

\begin{abstract}

  A statistical model of self-organization in a generic class of
  one-dimensional nonlinear Schr\"odinger (NLS) equations on a
  bounded interval is developed.  The main prediction of this model is
  that the statistically preferred state for such equations consists
  of a deterministic coherent structure coupled with fine-scale,
  random fluctuations, or radiation.  The model is derived from
  equilibrium statistical mechanics by using a mean--field
  approximation of the conserved Hamiltonian and particle number
  ($L^2$ norm squared) for finite-dimensional spectral truncations of
  the NLS dynamics.  The continuum limits of these approximated
  statistical equilibrium ensembles on finite-dimensional phase spaces
  are analyzed, holding the energy and particle number at fixed,
  finite values.  The analysis shows that the coherent structure
  minimizes total energy for a given value of particle number and
  hence is a solution to the NLS ground state equation, and that the
  remaining energy resides in Gaussian fluctuations equipartitioned
  over wavenumbers.  Some results of direct numerical integration of
  the NLS equation are included to validate empirically these
  properties of the most probable states for the statistical model.
  Moreover, a theoretical justification of the mean--field
  approximation is given, in which the approximate ensembles are shown
  to concentrate on the associated microcanonical ensemble in the
  continuum limit.

\vspace{1.5ex}


\vspace{1.5ex}
\noindent
Submitted to {\bf Physica D}

\end{abstract}

\section{Introduction}
\label{introsect}

The appearance of macroscopic organized states, or coherent
structures, in the midst of turbulent small-scale fluctuations is a
common feature of many fluid and plasma systems \cite{Hasegawa}.
Perhaps the most familiar example is the formation of large-scale
vortex structures in a turbulent, large Reynolds' number, two-dimensional
fluid \cite{McWilliams,SegreKida}.  A similar phenomenon occurs in
slightly dissipative magnetofluids in two and three dimensions, where
coherent structures emerge in the form of magnetic islands with flow
\cite{BW,KMT}.  In the present work, we shall be concerned with
another nonlinear partial differential equation whose solutions in the
long-time limit tend to form large-scale coherent structures while
simultaneously exhibiting intricate fluctuations on very fine spatial
scales.  Namely, we shall consider the class of one-dimensional wave
systems governed by a nonlinear Schr\"odinger equation of the form
\begin{equation} \label{nls1}
i \psi_t + \psi_{xx} + f(|\psi|^2)\psi = 0\,,
\end{equation}
for various nonlinearities $f$ chosen so that the dynamics is
nonintegrable and free of wave collapse.

Numerical simulations of the NLS equation (\ref{nls1}) in a bounded
spatial interval with either periodic or Dirichlet boundary conditions
demonstrate that, after a sufficiently long time, the field $\psi$
evolves into a state consisting of a time-periodic coherent structure
on the large spatial scales coupled with turbulent fluctuations, or
radiation, on the small scales \cite{DZPSY,ZPSY,JJ}.  For a focusing
nonlinearity $f$ ($f(a) > 0, f'(a) > 0$ for $a > 0$) the coherent
structure takes the form of a spatially localized, solitary wave.
This organization into a soliton-like state is observed from generic
initial conditions for the focusing NLS equation \cite{JJ}.  At
intermediate times a typical solution consists of a collection of the
soliton-like structures, which as time evolves undergo a succession of
collisions, or interactions.  In these collisions the larger of the
solitons increases in amplitude and smaller waves are radiated.  This
interaction of the solitons continues until eventually a single
soliton of large amplitude survives in a sea of small-scale turbulent
waves.  (Even though (\ref{nls1}) does not possess solitons in the
strict sense unless $f(|\psi|^2) = |\psi|^2$, it does have spatially
localized solitary wave solutions in the generic focusing case.  As a
matter of convenience, and in keeping with the terminology in
\cite{DZPSY,ZPSY}, we therefore simply refer to the coherent states
that emerge in the simulations in this case as solitons).  Figure
(1)  below illustrates this behavior for the particular saturated
focusing nonlinearity $f(|\psi|^2) = |\psi|^2/(1 + |\psi|^2 ) $.  For
a defocusing $f$, coherent structures also emerge and persist, but
they are less conspicuous since they are not spatially localized. The
reader is referred to \cite{JJ} for a detailed description of
numerical investigations of the long-time behavior of the system
(\ref{nls1}) for various nonlinearities $f$.

Theoretical arguments and numerical experiments suggest that these
coherent structures for the NLS dynamics (\ref{nls1}) correspond to 
ground states of the system.  That is, a coherent structure assumes
the form $\psi(x,t) = \phi(x) \exp(-i\lambda t)$, where $\phi(x)$ is a
solution of the ground state equation
\begin{equation} \label{gse}
\phi_{xx} + f(|\phi|^2)\phi + \lambda \phi = 0\,,
\end{equation}
In turn, these ground states $\phi$ are the minimizers of the 
Hamiltonian
\begin{equation} \label{ham1}
  H(\psi)= \frac{1}{2} \int_{\Omega} |\psi_x|^2\,dx - \frac{1}{2} 
  \int_{\Omega}F(|\psi|^2)\,dx \,
\end{equation}
subject to the constraint that the particle number
\begin{equation} \label{part1}
N(\psi) = \frac{1}{2} \int_{\Omega} |\psi|^2\,dx\
\end{equation}
be equal to given initial value $N^0$.  We shall
refer to the first and second terms in
(\ref{ham1}) as the {\em kinetic energy} and the {\em potential energy},
respectively. The potential $F$ is defined
by $F(a)= \int_0^a f(a') da'$. The parameter $\lambda$ arising in
(\ref{gse}) is the Lagrange multiplier in the variational principle:
minimize $H(\phi)$ subject to $N(\phi) = N^0$.  For
focusing nonlinearities $f$, the ground states $\phi$ are spatially
localized.  By contrast, for periodic boundary conditions and for
defocusing nonlinearities $f$, the ground states are spatially
uniform; this can be proved by a straightforward application of
Jenson's inequality when the potential $F$ is strictly concave.
Nevertheless, independent of the properties of the ground states, the
coherent structures are expected to be constrained minimizers of the
Hamiltonian given the particle number, whenever these minimizers
exist. 

Yankov and collaborators \cite{KY,DZPSY,ZPSY} have suggested that the
tendency of the solution of the NLS system in the focusing case to
approach a coherent soliton state coupled with small-scale radiation
can be understood within a statistical mechanics framework.  They
argue that the solutions to the ground state equation that realize the
minimum of the Hamiltonian subject to the constraint on the particle
number (\ref{gse}) are ``statistical attractors'' to which the
solutions of the NLS equation (\ref{nls1}) tend to relax.  In this
argument, the process which increases the amplitude of the solitons as
the number of solitons decreases is thought to be ``thermodynamically
advantageous'', in the sense that it increases the ``entropy'' of the
system.  This entropy seems to be directly related to the amount of
kinetic energy contained in the radiation, or the small-scale
fluctuations of the field $\psi$, so that the transference of kinetic
energy to the fluctuations is accompanied by an increase in the
entropy.  Similar ideas have been expounded by Pomeau \cite{Pomeau}
who has argued, based on weak turbulence theory, that the defocusing
cubic NLS equation ($f(|\psi|^2) = - |\psi|^2$) in a bounded
two--dimensional
 spatial
domain should exhibit the tendency to approach a long-time state
consisting of solution of the ground state equation (\ref{gse}) plus
small-scale radiation.

Our primary purpose in the present paper is to construct a statistical
equilibrium model of the coherent structures and the turbulent
fluctuations inherent in the long--time behavior of system
(\ref{nls1}).  With this model we seek to translate the various
intuitive ideas about the balance between order and disorder in the
NLS dynamics into explicit, verifiable calculations.  In particular,
we provide strong support to the notion set forth in
\cite{KY,DZPSY,ZPSY,Pomeau} that the ground states, which minimize the
energy for a given particle number, are statistically preferred
states.  In addition, we furnish a definite meaning to the concept
that random fluctuations absorb the remainder of the energy.

We derive our equilibrium statistical model from a mean--field
approximation of the conserved Hamiltonian $H$ and particle number $N$
for a finite--dimensional truncation of the NLS equation (\ref{nls1}).
This approximation relies on the fact that the fluctuations of $\psi$
about its mean, the coherent structure, become asymptotically small in
the continuum limit, provided that the mean energy and mean particle
number remain finite as the number of modes in the spectral
truncations goes to infinity.  In constructing the mean--field theory,
therefore, we choose that probability measure on the
finite--dimensional phase space which maximizes entropy subject to
approximated constraints on the mean energy and particle number.
Specifically, we approximate the means of the particle number and the
potential energy term in the Hamiltonian in terms of the mean state
$\langle \psi \rangle$.

This approach gives rise to a Gaussian model, which is accordingly
easy to analyze.  Thus, we find that to any initial value $N^0$ of the
particle number there corresponds a coherent structure, which is
precisely the mean state, or most probable state, for the statistical
model.  Moreover, we show that this mean state is indeed a solution to
the ground state equation that minimizes the Hamiltonian over all
states with the same particle number $N^0$.  Furthermore, we clarify
the role of fine-scale fluctuations by demonstrating that the gradient
field $\psi_x$ has finite variance at every point.  Thus, we see that
the difference between the initial energy $H^0$ and the energy of the
mean state, $H(\langle \psi \rangle$), resides entirely in the kinetic
energy of infinitesimally fine-scale fluctuations.  This result, which
makes precise the ideas of Yankov {\em et al.} \cite{DZPSY,ZPSY} and
Pomeau \cite{Pomeau}, explains how it is possible for an ergodic
solution to the Hamiltonian system (\ref{nls1}) to approach a solution
to the ground state equation (\ref{gse}) without violating the
conservation of energy or particle number.

In the present paper the principal justification for our mean--field
approximation is the evidence of extensive numerical simulations,
which show that the fluctuations in $\psi$ go to zero over longer and
longer time intervals \cite{DZPSY,ZPSY,JJ}.  Nevertheless, we also include an
{\em a posteriori} justification of this approximation by proving that,
in the continuum limit, the probability measures defining the
mean--field theory with approximately conserved quantities $H$ and $N$
concentrate about the phase space manifold on which $H=H^0$ and
$N=N^0$.  In other words, we establish a form of the equivalence of
ensembles, showing that our mean-field ensemble becomes equivalent in
a certain sense to the microcanonical ensemble, which is a measure
concentrated on the manifold $H=H^0, N=N^0$.  Besides connecting our
mean-field approach to the standard principles of equilibrium
statistical mechanics \cite{Balescu}, this result shows that the
mean--field theory is asymptotically exact in the continuum limit.
This property of the mean--field ensembles is a common feature of
turbulent continuum systems in which the mean values of conserved
quantities are fixed while the number of degrees of freedom goes to
infinity \cite{MWC,JT}.

These topics are taken up after a quick review of NLS theory and previous
work on invariant measures for NLS.

\section{The NLS Equation and its Conserved Quantities}
\label{NLSsect}

The NLS equation (\ref{nls1}) describes the slowly-varying envelope of
a wave train in a dispersive conservative system and so arises in many
branches of physics.  It models, for example, gravity waves on deep
water \cite{AS}, Langmuir waves in plasmas \cite{Pesceli}, pulse
propagation along optical fibers \cite{HK}, and self-induced
motion of vortex filaments \cite{Hasimoto,Majda}.  In particular, the cubic
NLS equation, which corresponds to the nonlinearity $f(|\psi|^2) = \pm
|\psi|^2,$ has garnered much attention in the mathematics and physics
literature.  On the whole real line, or on a periodic interval, the
cubic NLS equation is completely integrable via the inverse scattering
transform \cite{LM,ZS}. The NLS equation with any other nonlinearity
or boundary conditions, however, is not known to be integrable.

In the present article we are interested in the NLS equation
(\ref{nls1}) as a generic model of nonlinear wave turbulence.  For
this reason, we wish to consider nonlinearities $f$ and boundary
conditions such that the dynamics is well-posed for all time and yet
is nonintegrable.  For instance, a natural class of prototypes of this
kind consists of those NLS equations with bounded nonlinearities $f$
on a bounded spatial interval with Dirichlet boundary conditions.
Bounded nonlinearities arising in physical applications are often
called saturated nonlinearities, of which $f(|\psi|^2) = |\psi|^2/(1 +
|\psi|^2)$ and $f(|\psi|^2) = 1-\exp(-|\psi|^2)$ are examples.  These
amount to corrections to the focusing cubic nonlinearity for large
wave amplitudes, and they have been proposed as a models of nonlinear
self-focusing of laser beams \cite{AB}, propagation of nonlinear
optical waves through dielectric and metallic layered structures
\cite{MNF}, and self-focusing of cylindrical light beams in plasmas
due to the ponderomotive force \cite{max}.  Such nonlinearities
certainly satisfy the requirements set forth by Zhidkov to guarantee
well--posedness of the NLS equation (\ref{nls1}) in the space
$L^2$ for Dirichlet boundary conditions \cite{Zhidkov}.

In light of these properties, we shall adopt the NLS equation with
homogeneous Dirichlet boundary conditions on a bounded interval
$\Omega = [0, L]$ and with a (focusing or defocusing) bounded
nonlinearity $f$ as the generic system throughout our main
development.   Precisely, we impose on $f$ is the following 
{\em boundedness condition} 
\begin{equation} \label{boundcond}
\sup_{a \in R} \left ( |f(a)| + |(1 + a) f'(a)| \right ) \, < \, \infty \, . 
\end{equation} 
In Section 9 we indicate how our main results can be extended to a
natural class of unbounded nonlinearities.  Also, when it is
convenient for comparison with some numerical simulations, we can
instead use periodic boundary conditions.  The necessary
modifications, which are straightforward, are left to the reader.

The NLS equation (\ref{nls1}) may be cast in the Hamiltonian form 
\begin{displaymath}
 \frac{i}{2}\psi_t =\frac{\delta H}{\delta\psi^*} \,,
\end{displaymath}
where $\psi^*$ is the complex conjugate of the field $\psi$ and $H$ is
the Hamiltonian defined by (\ref{ham1}).  For our class of generic NLS
systems, the total energy $H$ and the particle number $N$, or $L^2$
norm squared, are the only known dynamical invariants for
(\ref{nls1}).  These invariants play the leading role in the
statistical equilibrium theory and the values $H^{0}$ and $N^{0}$,
derived from a given initial state, say, alone determine the
statistical ensemble.

For later use, we write the NLS equation (\ref{nls1}) as the following
coupled system of partial differential equations for the real and
imaginary components $u(x,t)$ and $v(x,t)$ of the complex field
$\psi(x,t)$:
\begin{equation} \label{nls2}
u_t + v_{xx} + f(u^2 + v^2)v = 0\,,\;\;\; v_t - u_{xx} - f(u^2+v^2)u=0\,.
\end{equation}  
Similarly, the Hamiltonian can be written in terms of
the fields $u$ and $v$ as
\begin{equation} \label{ham2}
  H(u,v)= K(u,v) \, + \,  \Theta (u,v) 
\end{equation}
with kinetic and potential energy terms
\[ 
 K(u,v) = \frac{1}{2} \int_{\Omega} ( u_x^2 + v_x^2)\, dx\,, \;\;\;\;\;\;\;\;
 \Theta(u,v) =-\frac{1}{2} \int_{\Omega} F(u^2+v^2) \, dx\,, 
\]
and the particle number is given by
\begin{equation} \label{part2}
  N(u,v)= \frac{1}{2} \int_{\Omega} (u^2+v^2)\, dx\,.
\end{equation}
In the same way, the ground state equation (\ref{gse}) for 
$\phi = u + i v$ can be expressed as 
the following coupled system of differential equations:
\begin{equation} \label{gse2}
u_{xx} + f(u^2 + v^2)u + \lambda u = 0\,,\;\;\;\; 
v_{xx} + f(u^2 + v^2)v + \lambda v = 0\,.
\end{equation}

\section{Invariant Measures for NLS}
\label{invsect}

Several authors have constructed invariant measures for the NLS
equation (\ref{nls1}).  Zhidkov \cite{Zhidkov} considered
nonlinearities $f$ satisfying the boundedness condition (\ref{boundcond})
and showed that the Gibbs measure
\begin{equation} \label{gibbs}
 P_{\beta}(dudv) = Z^{-1}\exp(-\beta H(u,v)) 
       \prod_{x \in \Omega} du(x) dv(x)\,,
\end{equation}
for fixed $\beta > 0$ is normalizable and has a rigorous
interpretation as an invariant measure for the NLS system on the phase
space $L^2$.  Similar results have been obtained by Bidegary
\cite{Bid}.  These results hold for both periodic and Dirichlet
boundary conditions.  However, $P_{\beta}$ as constructed by Zhidkov
fails to account for the invariance of the particle number
(\ref{part1}) under the NLS dynamics.

Earlier, Lebowitz {\em et al.} \cite{LRS} considered the problem of
constructing invariant measures for NLS on the periodic interval with
the focusing power law nonlinearities $f(|\psi|^2) = |\psi|^s$.  For
such nonlinearities the Hamiltonian $H$ is not bounded below, and
therefore the measure $P_{\beta} $ defined by (\ref{gibbs}) is not
normalizable.  As an alternative, Lebowitz {\em et al.} conditioned
the Gibbs measure on the particle number and studied the measures
\begin{equation} \label{mgibbs}
  P_{\beta,N^0}(dudv) = Z^{-1} \exp(-\beta H(u,v)) I(N(u,v) \le N^0) 
    \prod_{x \in \Omega} du(x) dv(x)\,,
\end{equation} 
where $I(A)$ is the characteristic function of the set $A$.  They
showed that $P_{\beta,N^0}$ is normalizable for $1\le s < 4$ for any
$N^0$, and for $s=4$ if $N^0$ is sufficiently small.  For $s>4$,
though, $P_{\beta,N^0}$ fails to be normalizable.  

The normalizability of the conditioned Gibbs ensemble (\ref{mgibbs})
is closely related to the well-posedness of the NLS equation as an
initial value problem \cite{LRS}.  Indeed, for the focusing power law
nonlinearities it can be shown that the NLS equation (\ref{nls1}) with
periodic or Dirichlet boundary conditions is well-posed, or free of
wave collapse, in the Sobolev space $H^1(\Omega)$ as long as $s < 4$,
while blow-up can occur from smooth initial conditions when $s \ge 4$
\cite{LRS,Bourgain,strauss}.  Bourgain \cite{Bourgain} proved that
under the conditions on $s$ and $N^0$ considered by Lebowitz {\em et
  al.} the measures $P_{\beta,N^0}$ are rigorously invariant under the
NLS dynamics and he thereby obtained a new global existence theorem
for the initial value problem with these nonlinearities. Analogous
results have also been established by McKean \cite{McKean}.    

A common feature of all these invariant measures for the NLS equation
is that for any fixed $\beta > 0$ the kinetic energy $(1/2) \int
(u_x^2 + v_x^2)\,dx$ is infinite with probability $1$, and hence the
Hamiltonian is also infinite with probability 1 \cite{LRS}.  This
difficulty occurs whenever classical Gibbs statistical mechanics is
applied to a system with infinitely many degrees of freedom and is
related to the well-known Jeans ultraviolet catastrophe
\cite{Balescu}.  Consequently, these ensembles are appropriate when
the typical state of the continuum system is such that each of its
modes contains a finite energy.  They are not appropriate, on the
other hand, to situations in which a typical state of the system is
realized by an ergodic evolution from an ensemble of initial
conditions having finite energy $H^0$ and particle number $N^0$, both
of which partition amongst the many modes of the system.
  
Since the purpose of the current paper is to construct ensembles which
provide meaningful predictions about the long--time behavior of
solutions of the NLS equation (\ref{nls1}), we are interested in
statistical equilibrium ensembles which, in the continuum limit, have
finite mean energy and mean particle number.  In view of this goal, we
scale the inverse temperature parameter $\beta$ in the Gibbs measure
with the number of degrees of freedom $n$ and thereby maintain the
mean energy at a finite value as $n \rightarrow \infty$.  The
continuum limit we obtain under this scaling is degenerate in certain
ways, but this is hardly surprising given that the known invariant
measures for NLS are supported on fields with infinite kinetic energy.
On the other hand, the analysis of the scaled continuum limit is
greatly simplified by the fact that the random fluctuations of $(u,v)$
at each point tend to zero.  This fortuitous property implies that a
natural mean--field approximation is asymptotically exact.  Moreover,
the mean-field ensemble  concentrates on the microcanonical
ensemble for the invariants $H$ and $N$ in the scaled continuum limit.
For this reason, we can avoid the canonical ensemble constructed from
$H$ and $N$, which fails to be normalizable for focusing
nonlinearities because the linear combination $H - \lambda N$ always
has a direction in which it goes to $-\infty$, as we shall demonstrate
in Section 5.

\vspace{1.5ex}

\section{Discretization of NLS}
\label{discretesect}

We now introduce a finite--dimensional approximation of the NLS
equation (\ref{nls1}) with homogeneous Dirichlet boundary conditions
on the interval $\Omega = [0,L]$.  For this purpose we shall use a
standard spectral truncation, even though many other approximation
schemes could be invoked.  Let $e_k(x)= \sqrt{2/L} \sin
(\sqrt{\lambda_k} x)$ and $\lambda_k= ( k \pi /L)^2, k=1,2,\ldots$, be
the eigenfunctions and eigenvalues of the operator $-
\frac{d^2}{dx^2}$ on $\Omega$ with homogeneous Dirichlet boundary
conditions.  The eigenfunctions $e_k$ 
form an orthonormal basis for the real space $L^2 (\Omega) $. 
For given $n$, let $X_n = \mbox{span } \{e_1, \ldots, e_n \} \subset
L^2$, define $\un$ and $\vn$ in $X_{n}$ by
\begin{equation} \label{unvn}
  u^{(n)}(x,t) = \sum_{k=1}^n u_k(t) e_k(x) \,,\;\; 
  v^{(n)}(x,t) = \sum_{k=1}^n v_k(t) e_k(x) \,,
\end{equation}
and set $\psi^{(n)}(x,t) = u^{(n)}(x,t) + i v^{(n)}(x,t)$.
We consider the following evolution equation for $\psi^{(n)}$:
\begin{equation} \label{spectrunc1}
  i \psi^{(n)}_t + \psi^{(n)}_{xx} + P^{(n)} \left (
f(|\psi^{(n)}|^2)\psi^{(n)}
  \right ) = 0\,,
\end{equation}
where $P^{(n)}$ denotes the orthogonal projection from $L^{2}(\Omega)$
onto $X_n$.  Equation (\ref{spectrunc1}) is clearly a spectral
truncation of the NLS equation (\ref{nls1}).  It is equivalent to the
following coupled system of ordinary differential equations for the
Fourier coefficients $u_k$ and $v_k, k= 1,\ldots,n$:
\begin{equation}\label{spectrunc2}
  \frac{du_k}{dt} - \lambda_k v_k + 
  \int_{\Omega} f( (u^{(n)})^2 + (v^{(n)})^2) v^{(n)} e_{k}\,dx = 0\,,
\end{equation}
\begin{equation}\label{spectrunc3}
  \frac{dv_k}{dt} + \lambda_k u_k - 
  \int_{\Omega} f( (u^{(n)})^2 + (v^{(n)})^2) u^{(n)} e_{k}\,dx = 0,
\end{equation}
where we have used the fact that $(e_k)_{xx}= - \lambda_k e_k$.

Whether viewed as the evolution of $(\un,\vn)$ on $X_{n}\times X_{n}$
or in terms of the Fourier coefficients $(u_{k},v_{k})$ on $R^{2n}$,
it can be shown \cite{Zhidkov,Bid} that this finite--dimensional
dynamical system has Hamiltonian structure, with Hamiltonian
$H_n=K_n+\Theta_n$, where
\begin{equation}\label{kn}
  K_n(\un,\vn) = \frac{1}{2} \int_{\Omega} [(u^{(n)}_{x})^{2} +
                                          (v^{(n)}_{x})^{2}] \,dx
             = \frac{1}{2} \sum_{k=1}^n \lambda_k (u_k^2 + v_k^2)\,,
\end{equation}
is the kinetic energy, and
\begin{equation} \label{tn}
  \Theta_n(\un,\vn) = -\frac{1}{2}\int_{\Omega} F((u^{(n)})^2 + 
  (v^{(n)})^2)\, dx\,,
\end{equation}
is the potential energy. The functionals $H_n, K_n$ and $\Theta_n$ are
just the restrictions to $X_n \times X_n$ of the functionals $H, K$
and $\Theta$, which are defined by (\ref{ham2}).  The Hamiltonian
$H_n$ is, of course, an invariant of the dynamics.  The particle
number
\begin{equation}\label{dispn}
  N_n(\un,\vn) = \frac{1}{2} \int_{\Omega} [(u^{(n)})^{2} +
  (v^{(n)})^{2}] \,dx = \frac{1}{2} \sum_{k=1}^n (u_k^2 + v_k^2)
\end{equation}
is also conserved by this dynamics, as may be verified by direct
calculation.  Also, $N_n$ is the restriction of the functional $N$,
which is defined by (\ref{part2}), to the space $X_n \times X_n$.

The Hamiltonian system (\ref{spectrunc2})-(\ref{spectrunc3}) as a 
dynamical system on $R^{2n}$ satisfies the
Liouville property \cite{Bid}. In other words, the Lebesgue measure
$\prod_{k=1}^n du_k dv_k$ on $R^{2n}$ is invariant under the phase
flow for this dynamics.

\section{Mean-Field Ensembles for NLS}
\label{idealsect}

We now proceed to construct  a statistical model that describes the
long--time behavior of the truncated NLS system
(\ref{spectrunc2})-(\ref{spectrunc3}). According to the fundamental
principles of equilibrium statistical mechanics, the Liouville
property and the ergodicity of the dynamics provide the usual starting
point for such a description, and the microcanonical ensemble is the
appropriate statistical distribution (that is, probability measure on
phase space) with which to calculate averages for an isolated 
Hamiltonian system \cite{Balescu,PB}.  In the specific system under
consideration, the natural canonical random variables are $u_{k}$ and
$v_{k}$, which comprise the phase space $R^{2n}$.  The random fields
$\un(x) = \sum u_{k} e_{k}(x)$, $\vn(x) = \sum v_{k} e_{k}(x)$, and
$\psin(x)=\un(x)+i \vn(x)$, as well as the functions
$H_{n}=H(\un,\vn)$ and $N_{n}=N(\un,\vn)$, are determined by these
canonical variables.  The microcanonical ensemble for the truncated
NLS system therefore takes the form
\begin{equation} \label{microcan}
  P_{H^0,N^0}(dudv) \;=\; W^{-1} \delta (H_n-H^0) \, \delta (N_n-N^0)
  \, \prod_{k=1}^n du_k dv_k\, ,
\end{equation} 
where $W=W(H^0,N^0)$ is the structure function or normalizing factor.
Under the ergodic hypothesis, expectations with respect to this
ensemble equal long-time averages over the spectrally-truncated
dynamics from initial conditions with prescribed, finite values $H^0$
and $N^0$ of the invariants $H_n$ and $N_n$. 

While the microcanonical ensemble is a well-defined invariant measure
for the spectrally-truncated dynamics, it is a cumbersome to analyze,
and the calculation of ensemble averages for finite $n$ is not feasible.
In statistical mechanics the usual procedure to overcome this
difficulty is to introduce another invariant measure, the
canonical ensemble, and prove that in the limit as $n \rightarrow
\infty$, it becomes equivalent to the microcanonical ensemble
\cite{Balescu,PB}.  For the NLS system the canonical Gibbs ensemble is
\begin{equation} \label{can}
P_{\beta,\lambda} (dudv) \;=\; Z^{-1} \exp(-\beta [H_{n} - \lambda N_{n}]) 
\, \prod_{k=1}^n du_k dv_k,
\end{equation}
where $\beta$ and $\lambda$ are chosen such that the averages of $H_n$
and $N_n$ with respect to $P_{\beta,\lambda}$ equal $H^0$ and $N^0$,
respectively, and $Z = Z(\beta,\lambda)$ is the partition function or
normalizing factor.  Unfortunately, for general focusing
nonlinearities $f$, even those satisfying the boundedness condition
(\ref{boundcond}), this canonical measure is not normalizable.  To see
this degeneracy, let us consider a nonlinearity $f$ that is both
positive and strictly increasing and an associated ground state $\phi$
that is a nontrivial solution to (\ref{gse}).  Then, for any positive
scale factor $\sigma$, a straightforward calculation yields the
identity
\begin{eqnarray*}
  (H - \lambda N) ( \sigma \phi) &=& \frac{\sigma^2}{2} \int_{\Omega}[
  |\phi_x|^2 \, - \, \lambda | \phi|^2 ] \, dx \,-\, \frac{1}{2}
  \int_{\Omega} F( \sigma^2 |\phi|^2) \, dx \\ &=& - \frac{1}{2}
  \int_{\Omega} dx \, \int_0^{ \sigma^2 |\phi|^2} [ f(a) -
  f(|\phi|^2) ] \, da \; .
\end{eqnarray*}
It follows from this expression that, as $\sigma$ increases to
infinity, the value of $H - \lambda N$ at $\sigma \phi$ tends to $-
\infty$.  The same reasoning applies to the discretized system.
Consequently, the function $H_n - \lambda N_n$ in the Gibbs weight has
a direction in which is goes to $- \infty$, resulting in the
divergence of the partition function $Z$ in (\ref{can}). In essence,
this divergence reflects the fact that such a ground state $\phi$ is a
critical point, but not a minimizer, of $H - \lambda N$.  Locally, the
first variation $\delta H - \lambda \delta N$ vanishes at $\phi$, but
the second variation $\delta^2 H - \lambda \delta^2 N$ is positive
only on those variations $\delta \phi$ for which $\delta N = 0$.  When
$\delta \phi = (1 - \sigma) \phi$ for $\sigma$ near $1$, and thus
$\delta N \ne 0$, the corresponding second variation of $H - \lambda
N$ is negative.  The above calculation shows that this local behavior
extends to a global degeneracy for a wide class of nonlinearities. 
As Lebowitz {\it et al.} discuss in
\cite{LRS}, this defect of the canonical ensemble is even more severe
when the nonlinearity $f$ is unbounded.

These considerations compel us to find another way to approximate the
microcanonical ensemble and thus to derive a tractable statistical
equilibrium model.  First, we note that since the microcanonical
measure (\ref{microcan}) depends on $\un$ and $\vn$ through $H_{n}$
and $N_{n}$ only, it is invariant with respect to multiplicative
constants of the form $e^{i \theta}$. In view of this phase
invariance, we can factor the random field $\psin$ in the form $e^{i
  \theta} \phin$, where $\theta$ is uniformly distributed on
$[0,2\pi]$ and is independent of $\phin$.  
In what follows, we shall write
$\phin=\un+i\vn$ with the phase normalization
\[
\mbox{ arg } \, \int_{\Omega} \phin(x) \, dx \;=\;0\,,
\]
and consider the microcanonical distribution conditioned by this
constraint.  It is this conditional distribution that we seek to
approximate with a mean-field approximation.                          
In constructing the approximation, we can therefore
describe distributions for $\phin$ that need not possess invariance
under a change of phase, with the understanding that we
can recover the random field $\psin$ by setting $\psin=e^{i \theta}
\phin$.

The key to the mean-field construction is supplied by direct numerical
simulations of the evolving microstates $\psi$ governed by
(\ref{nls1}) \cite{ZPSY,JJ}.
Since the particle number and the Hamiltonian are well conserved
in these simulations, we may think of these numerical experiments as 
realizations of the microcanonical ensemble.  The simulations clearly
demonstrate that the amplitude of fluctuations of the random field
$\phi$ (the phase normalized field associated with $\psi$) at each
point $x \in \Omega$ become small in the long-time limit, and that
they contribute little to the conserved $L^2$ norm of $\phi$
\cite{ZPSY,JJ} (also, see Figure 1 below).  
Moreover, this effect becomes more apparent as the
spatial resolution of the numerical simulations is improved \cite{JJ}.
In particular, these fluctuations become negligible compared to the
magnitude of the coherent structure that emerges in the mean field
$\phi$, which in the focusing cases takes the form of a single,
localized soliton.  On the basis of these properties of the simulated
microstates, we shall adopt the hypothesis that, in the limit of
infinite resolution, the fluctuations of $\phi$ are infinitesimal.
Precisely, we shall make the following {\em vanishing of fluctuations
  hypothesis}: 

\vspace*{.2in}
\noindent
With respect to the phase-invariance conditioned microcanonical
ensemble on $2n$-dimensional phase-space, there holds 
\begin{equation} \label{hyp1}
  \lim_{n \rightarrow \infty} \int_{\Omega} [\mbox{Var}( \un(x)) +
  \mbox{Var}(\vn(x))] \, dx \;=\; \lim_{n \rightarrow \infty}
  \sum_{k=1}^n [ \mbox{Var}(u_{k}) + \mbox{Var}(v_{k})] \;=\, 0 \, ,
\end{equation}
where Var denotes the variance of the indicated random variable with
respect to that ensemble.  The first equality follows from the identity
$ \int_{\Omega} [ \mbox{Var}(\un(x)) + \mbox{Var}(\vn(x)) ] \, dx = 
\sum_{k=1}^n [ \mbox{Var}(u_k) + \mbox{Var}(v_k) ]$.

\vspace{1.5ex}

Our strategy is to use the vanishing of fluctuations hypothesis to
build a mean-field ensemble that is equivalent in an appropriate
sense to the microcanonical
ensemble (\ref{microcan}) in the continuum limit as $n \rightarrow
\infty$, but is easy to analyze for finite $n$.  Rather than attempt
to verify this hypothesis {\it a priori}, we shall show {\it a
  posteriori} that the resulting mean-field ensemble concentrates on
the microcanonical manifold at fixed $H^0$ and $N^0$.   The analysis
of the fundamental microcanonical
ensemble itself will be taken up in a subsequent publication.

We note however that the numerical simulations of the NLS system which
support the hypothesis (\ref{hyp1}) also strongly indicate that the
fluctuations of the gradients $u^{(n)}_x, v^{(n)}_x$ are not
negligible in the limit of infinite resolution.  Figure 2
illustrates the typical long-time behavior of $|\psi_x|^2$ for the
particular nonlinearity $f(|\psi|^2)= |\psi|^2/(1 + |\psi|^2)$, in
which finite fluctuations in the gradient $\psi_x$ persist.  This
property of the microcanonical ensemble is compatible with the
vanishing of fluctuations hypothesis, as can be seen from the
straightforward identity
\begin{equation} \label{gradfluct}
    \int_{\Omega} [\mbox{Var}(u^{(n)}_{x}(x)) +
\mbox{Var}(v^{(n)}_{x}(x))] dx
  = \sum_{k=1}^n \lambda_{k} [ \mbox{Var}(u_{k}) + \mbox{Var}(v_{k})]\,.
\end{equation}
In fact, as we shall see, this quantity remains finite in the
continuum limit according to our statistical theory, and it 
 represents the portion of kinetic energy absorbed by
infinitesimally fine-scale fluctuations.  

We now proceed to construct a mean-field ensemble for each $n$ based
on approximations derived from the hypothesis (\ref{hyp1}).  This
ensemble is determined by a probability density $\rhon$ on $R^{2n}$
with respect to Lebesgue measure $\prod_{k=1}^n du_k dv_k$.  (Unlike
the microcanonical ensembles, the mean-field ensembles are taken to be
absolutely continuous with respect to Lebesgue measure for each $n$.)  In
order to define that $\rhon$ which governs the mean-field theory, we
appeal to standard statistical-mechanical and information-theoretic
principles \cite{Jaynes,Balescu,PB} and choose  $\rhon$ so that it
maximizes the Gibbs--Boltzmann entropy functional
\begin{equation} \label{entropy}
  S(\rho) = -\int_{R^{2n}} \rho(u_1,\ldots, u_n, v_1, \ldots, v_n)
  \log \rho(u_1,\ldots, u_n, v_1, \ldots, v_n)\prod_{k=1}^n du_k dv_k\,,
\end{equation}
subject to some appropriate constraints.  The entropy $S$ has the
well-known interpretation as a measure of the number of {\em
  microstates} $(u, v)$ corresponding to the {\em macrostate} $\rho$
\cite{Balescu}.  The form of the entropy as a functional of $\rho$ is
dictated by the Liouville property of the dynamics
(\ref{spectrunc2})--(\ref{spectrunc3}), which requires that the
entropy be relative to the invariant measure
$\prod_{k=1}^n du_k dv_k$. Alternatively, from the information
theoretic point of view, maximizing the entropy $S$ amounts to finding
the least biased distribution on $2n$-dimensional phase space
compatible with the constraints and the uniform prior distribution
$\prod_{k=1}^n du_k dv_k$ \cite{Jaynes}.   

The constraints imposed in the Maximum Entropy Principle (MEP) are the
crucial ingredient in the determination of the ensemble $\rhon$.  As
is well-known, the microcanonical ensemble is produced when they are
taken to be the exact constraints $H_n=H^0$ and $N_n=N^0$; the canonical
ensemble, which however is ill-defined in the focusing case,
corresponds to average constraints $\langle H_n \rangle = H^0$ and
$\langle N_n \rangle = N^0$.  Throughout our discussion of (MEP) the
angle brackets $\langle \cdot \rangle$ denote expectation with respect
to the admissible density $\rho$ in (MEP).  For our mean-field theory
we seek constraints that are intermediate between those giving the
microcanonical and the canonical ensembles.  To this end we invoke the
vanishing of fluctuations hypothesis (\ref{hyp1}) and impose the
following constraints:
\begin{equation}\label{mfc1}
 \frac{1}{2} \sum_{k=1}^n (\langle u_k \rangle^2
+ \langle v_k \rangle^2) = N^0\,.
\end{equation}
\begin{equation}\label{mfc2}
 \frac{1}{2} \sum_{k=1}^n \lambda_k (\langle
u_k^2 \rangle + \langle v_k^2 \rangle) 
-\frac{1}{2} \int_{\Omega} F(\langle u^{(n)} \rangle^2 + \langle v^{(n)}
\rangle^2)\,dx = H^0 ,
\end{equation}
which we shall refer to as the {\em mean--field constraints} for
obvious reasons.  We note that these constraints involve only the
first and second moments of the fields $\un$ and $\vn$ with respect to
$\rho$.

We can motivate this choice of constraints in (MEP) by approximating
the values of the conserved quantities $H_n$ and $N_n$ in terms of
means with respect to the microcanonical ensemble.  An immediate
implication of (\ref{hyp1}) applied to the definition of $N_n$ given
in (\ref{dispn}) is that
\begin{eqnarray}\label{Napprox}
  N^0 &=& \frac{1}{2} \int_{\Omega} (\unxmean^{2} +
  \vnxmean^{2}) \,dx \;+\; \frac{1}{2} \int_{\Omega} [\mbox{Var}(\un(x)) +
  \mbox{Var}(\vn(x))] \, dx \nonumber \\   &=&
  N_n(\unmean, \vnmean) \;+\; o(1)    
\end{eqnarray}
as $n \rightarrow \infty$.  When we drop the error term in this
approximation at finite $n$ and we replace expectation with respect to
the microcanonical distribution by expectation with respect to an
admissible density $\rho$, we obtain the mean-field constraint
(\ref{mfc1}).

Similarly, we can obtain the mean-field constraint (\ref{mfc2}) from
the vanishing of fluctuations hypothesis (\ref{hyp1}) by analyzing $
H_n$ with respect to the microcanonical distribution for large $n$.
The kinetic energy (\ref{kn}) in $ H^0 = \langle H_n \rangle$ is
retained exactly as $\langle K_n \rangle$.  On the other hand, the
potential energy (\ref{tn}) is approximated by expanding $F$ about the
mean $(\langle u^{(n)} \rangle, \langle v^{(n)} \rangle)$, which
yields
\begin{eqnarray} \label{tnexp}
  \Theta_n(\un,\vn) & = & -\frac{1}{2} \int_{\Omega} F(\langle u^{(n)}
  \rangle^2 + \langle v^{(n)} \rangle^2)\,dx \nonumber \\ &-&
  \int_{\Omega} f(\langle u^{(n)} \rangle^2 + \langle v^{(n)}
  \rangle^2)\left ( \langle u^{(n)} \rangle (u^{(n)} - \langle u^{(n)}
    \rangle ) + \langle v^{(n)} \rangle (v^{(n)} - \langle v^{(n)}
    \rangle ) \right )\, dx \nonumber \\ & - & \frac{1}{4}
  \int_{\Omega} \left ( \begin{array}{c} (u^{(n)} - \langle u^{(n)}
      \rangle) \\ (v^{(n)} - \langle v^{(n)} \rangle) \end{array}
  \right )^{T} J(\tilde{u}^{(n)}, \tilde{v}^{(n)}) \left (
    \begin{array}{c} (u^{(n)} - \langle u^{(n)} \rangle) \\ (v^{(n)} -
      \langle v^{(n)} \rangle) \end{array} \right )\,dx\,,
\end{eqnarray}
where 
\begin{equation} \label{matrix}
  J(u,v) = \left ( \begin{array}{cc} 2f(u^2 + v^2) + 4u^2f'(u^2 + v^2)
      & 4uv f'(u^2 + v^2)\\ 4uv f'(u^2 + v^2) & 2f(u^2 + v^2) + 4v^2
      f'(u^2 + v^2)
                      \end{array} \right )
\end{equation}
is the matrix of second partial derivatives of $F(u^2 + v^2)$ with
respect to $u$ and $v$ and $(\tilde{u}^{(n)}, \tilde{v}^{(n)})$ lies
between $(u^{(n)}, v^{(n)})$ and $(\langle u^{(n)} \rangle, \langle
v^{(n)} \rangle)$.  Because we are considering nonlinearities $f$
satisfying (\ref{boundcond}) each element of the matrix
$J(\tilde{u}^{(n)}, \tilde{v}^{(n)})$ is bounded over the domain
$\Omega$ independently of $n$.  Therefore, taking the expectation on
both sides of equation (\ref{tnexp}) and noting that the second term
on the right hand side of this equation has zero mean, we obtain
\begin{displaymath}
\langle \Theta_n \rangle = -\frac{1}{2} \int_{\Omega} F(\langle u^{(n)}
\rangle^2 + \langle v^{(n)} \rangle^2)\,dx
+ R_n\,,
\end{displaymath}
where the remainder $R_n$ satisfies 
\begin{eqnarray*}
  |R_n| & \le & C \left \langle \int_{\Omega} [(u^{(n)} - \langle u^{(n)}
  \rangle)^2 + (v^{(n)} -\langle v^{(n)} \rangle)^2
  ]\,dx \right \rangle  \\ 
  & = & C \int_{\Omega} [\mbox{Var}(\un(x)) + \mbox{Var}(\vn(x))] \, dx 
\end{eqnarray*}
for some constant $C$ independent of $n$.  The vanishing of
fluctuations hypothesis (\ref{hyp1}) thus implies that $R_n \rightarrow 0$
as $n \rightarrow \infty$, and hence that 
\begin{equation} \label{Happrox}
  H^0 \;=\; \frac{1}{2} \sum_{k=1}^n \lambda_k
  (\langle u_k^2 \rangle + \langle v_k^2 \rangle) -\frac{1}{2}
  \int_{\Omega} F(\langle u^{(n)} \rangle^2 + \langle v^{(n)}
  \rangle^2)\,dx \;+\; o(1).  
\end{equation}
The mean-field constraint (\ref{mfc2}) results from neglecting the error
term in this expression for finite $n$ and replacing microcanonical
expectations by expectations with respect to $\rho$.   

In summary, the mean-field theory is defined by a probability density
$\rhon$ on $R^{2n}$ that maximizes the entropy $S$ given in
(\ref{entropy}) subject to the constraints (\ref{mfc1}) and
(\ref{mfc2}).  These mean-field ensembles, which solve the governing
(MEP), are analyzed in detail in the subsequent sections.  Here we
merely note that they enjoy some properties not shared by either the
microcanonical or the canonical ensembles.  First, unlike the
microcanonical ensemble, the mean-field ensembles are 
analytically tractable because the densities $\rho^{(n)}$ 
are Gaussian.  This
desirable property is a simple consequence of the fact that the
constraints (\ref{mfc1}) and (\ref{mfc2}) involve only the first and
second moments of $\rho^{(n)}$.  Second, in contrast to the canonical
ensemble, the mean-field ensemble exists in both the focusing and
defocusing cases.  This crucial property depends on the fact that
fluctuations in $N_n$ are suppressed in the mean-field ensemble.   

However, because of the approximations made in deriving the
constraints (\ref{mfc1}) and (\ref{mfc2}), the mean-field ensemble
$\rho^{(n)}$ is not an invariant measure for the truncated NLS
dynamics (\ref{spectrunc2})-(\ref{spectrunc3}) at finite $n$, even
with the random phase $e^{i \theta}$  included.  Nevertheless, it
becomes consistent with the fundamental microcanonical ensemble in the
limit as $n \rightarrow \infty$, in the sense that
$ \langle N_n \rangle \rightarrow N^0$,
$\langle H_n \rangle \rightarrow H^0$, 
and the variances of $N_n$ and $H_n$ converge to 0 .  These
properties, which are proved in Section 8, imply that the ensembles
$\rhon$ concentrate about the phase space manifold $H_n=H^0, N_n=N^0$,
and thereby justify the approximations made in deriving the mean-field
theory.

\section{Equilibrium States for the Mean-Field Theory}
\label{approxsect}

We now proceed to calculate and analyze the solutions $\rhon$ to the
maximum
entropy principle
(MEP).  First, we calculate the density $\rho^{(n)}$ which maximizes
entropy subject to the mean-field constraints
(\ref{mfc1})-(\ref{mfc2}) by the method of Lagrange multipliers.  If
we denote the left-hand sides of (\ref{mfc1}) and (\ref{mfc2}) by
$\tilde{N}_{n}$ and $\tilde{H}_{n}$, then we have
\begin{equation}\label{lmr}
  \delta S =  \mun \delta \tilde{N}_n + \betan \delta \tilde{H}_n\,,
\end{equation}
where $\delta$ denotes variation with respect to the density variable
$\rho$ and $\mun$ and $\betan$ are the Lagrange multipliers to
enforce the constraints (\ref{mfc1}) and (\ref{mfc2}) on $\tilde{N_n}$
and $\tilde{H_n}$.  The variation of $S$ is 
\begin{equation}\label{delS}
  \delta S = - \int_{R^{2n}} \log (\rho) \delta \rho \prod_{k=1}^n
  du_k dv_k \,,
\end{equation}
for variations $\delta \rho$ satisfying $\int_{R^{2n}} \delta \rho
\prod_{k=1}^n du_k dv_k =0$.  Similarly, the variations of
$\tilde{N}_n$ and $\tilde{H}_n$ are 
\begin{equation} \label{delN}
  \delta \tilde{N}_n 
  = \sum_{k=1}^n \int_{R^{2n}} ( \langle u_k \rangle u_k +
  \langle v_k \rangle v_k)\, \delta \rho
  \prod_{k=1}^n du_k dv_k\,,
\end{equation} 
and
\begin{eqnarray} \label{delH}
  \lefteqn{ \delta \tilde{H}_n = 
  \sum_{k=1}^n \frac{1}{2}\int_{R^{2n}} \lambda_k ( u_k^2 + v_k^2)\, 
  \delta \rho  \prod_{k=1}^n du_k dv_k  }  \nonumber \\ 
   &-&  \int_{R^{2n}} \left ( \int_{\Omega} f(\langle u^{(n)} \rangle^2 +
  \langle v^{(n)} \rangle^2)
  \left [\langle u^{(n)} \rangle \un
  + \langle v^{(n)} \rangle \vn \right ]\, dx \right) \delta
  \rho \prod_{k=1}^n du_k dv_k\, 
\end{eqnarray}
using the relation $F' = f$.  Equating the coefficients of $\delta
\rho$ in (\ref{lmr}) and recalling that $\un = \sum_{k=1}^{n} u_{k}
e_{k}$ and $\vn = \sum_{k=1}^{n} v_{k} e_{k}$, we discover after some
algebraic manipulations that the entropy--maximizing density
$\rho^{(n)}$ has the form
\begin{equation}\label{rho1}
  \rho^{(n)}(u_1, \ldots, u_n, v_1, \ldots, v_n) = \prod_{k=1}^n
\rho_k(u_k,
  v_k)\,,
\end{equation}
where, for $k=1,\ldots, n$, 
\begin{equation}\label{rho2}
  \rho_k(u_k,v_k) \,=\, \frac{\betan \lambda_k}{2 \pi} \exp \left \{
    -\frac{\betan \lambda_k}{2} \left [ (u_{k} - \langle u_k \rangle
      )^2 + (v_{k} - \langle v_k \rangle )^2 \right ] \right\}\,,
\end{equation}
with
\begin{equation}\label{rho3}
  \langle u_k \rangle = \frac{1}{\lambda_k} \int_{\Omega} f(\langle
  u^{(n)}(x) \rangle^2 + \langle v^{(n)}(x) \rangle^2) \langle
  u^{(n)}(x) \rangle e_k(x) \, dx -\frac{\mun}{\betan \lambda_k}\langle
  u_k \rangle\,,
\end{equation}
\begin{equation}\label{rho4}
  \langle v_k \rangle = \frac{1}{\lambda_k} \int_{\Omega} f(\langle
  u^{(n)}(x) \rangle^2 + \langle v^{(n)}(x) \rangle^2) \langle
  v^{(n)}(x) \rangle e_k(x) \, dx -\frac{\mun}{\betan \lambda_k}\langle
  v_k \rangle\,.
\end{equation}

We see immediately that $u_{1}, \ldots, u_{n}, v_{1}, \ldots, v_{n}$
are mutually independent Gaussian random variables with means 
satisfying (\ref{rho3})--(\ref{rho4}), and variances
\begin{equation}\label{vark}
  \mbox{Var}(u_k) = \mbox{Var}(v_k) = \frac{1}{\betan \lambda_k}\,.
\end{equation}
The multiplier $\betan$ is therefore necessarily positive.  The
equations (\ref{rho3})--(\ref{rho4}) for the means $\langle u_k
\rangle$ and $\langle v_k \rangle$ can be written as an equivalent
complex equation for the mean field $\langle \phi^{(n)} \rangle =
\langle u^{(n)}\rangle + i \langle v^{(n)} \rangle$:
\begin{equation} \label{mfeqn}
  \langle \phi^{(n)} \rangle_{xx} \,+\, P^{(n)} \left (f(|\langle
    \phi^{(n)} \rangle |^2) \langle \phi^{(n)} \rangle \right ) \,+\,
  \lambdan \langle \phi^{(n)}\rangle \, = \, 0 \, ,
\end{equation}
where we introduce the real parameter $ \lambdan = -\mun/\betan$.
Recalling that $P^{(n)}$ is the projection onto the span $X_{n}$ of
the first $n$ eigenfunctions $e_1, \ldots, e_n$, we recognize this
equation as the spectral truncation of the ground state equation
(\ref{gse}) for the continuous NLS system (\ref{nls1}).  In other
words, the mean field $\langle \phin \rangle$ at each finite $n$ is a
critical point of the functional $ H_n - \lambdan N_n = 0$, by virtue
of (\ref{mfeqn}).  Thus, we draw the important conclusion that the
mean field predicted by the  statistical equilibrium theory is a solution
of the ground
state equation for the NLS system.

Since the maximum entropy distribution $\rho^{(n)}$ satisfies the
mean--field Hamiltonian constraint (\ref{mfc2}), a direct calculation
reveals that
\begin{eqnarray} \label{hameqn1}
  H^0 & = & \frac{1}{2}\sum_{k=1}^n \lambda_k ( \mbox{Var }u_k +
  \mbox{Var }v_k ) + \frac{1}{2}\sum_{k=1}^n \lambda_k ( \langle u_k
  \rangle^2 + \langle v_k \rangle^2 ) - \frac{1}{2} \int_{\Omega} F(
  \langle u^{(n)}\rangle^2 + \langle v^{(n)}\rangle^2)\, dx \nonumber
  \\ & = & \frac{n}{\betan} + H_n ( \langle u^{(n)} \rangle, \langle
  v^{(n)} \rangle)\,,
\end{eqnarray}
where (\ref{vark}) is invoked to obtain the second equality.  In this
expression the term $n/\betan$ in represents the contribution of
fluctuations to the energy, while $ H_n ( \langle u^{(n)}
\rangle,\langle v^{(n)} \rangle)$ is the energy of the mean state, or
coherent structure.  We can rearrange (\ref{hameqn1}) to get the
following expression for $\betan$ in terms of the number of modes $n$
and the energy of the mean state:
\begin{equation} \label{betan}
\betan = \frac{n}{H^0 - H_n ( \langle u^{(n)} \rangle,\langle v^{(n)}
\rangle)}\,.
\end{equation}

Now we come to the following central result, which enables us to
characterize completely the mean--field statistical ensembles
$\rho^{(n)}$.   

\vspace{1.5ex}
\noindent
{\bf Theorem 1.} The mean field pair $(\langle u^{(n)} \rangle,
\langle v^{(n)}\rangle)$ corresponding to any solution $\rho^{(n)}$ of
the constrained variational principle (MEP) is an absolute minimizer
of the Hamiltonian $H_n$ over all $(u^{(n)},v^{(n)}) \in X_n \times
X_n$ which satisfy the particle number constraint $N_n
(u^{(n)},v^{(n)}) = N^0$.

\vspace{1.5ex}
\noindent
{\bf Proof:} To prove this assertion, we calculate the entropy of a
solution $\rho^{(n)}$ of (MEP).  Referring to equations
(\ref{rho1})--(\ref{vark}), which define any solution of (MEP) we
immediately obtain

\begin{eqnarray*}
 S(\rho^{(n)})  & = & -\int_{R^{2n}} \rho^{(n)}
\log \rho^{(n)}\prod_{k=1}^n du_k dv_k
\\
& = & -\sum_{k=1}^n \int_{R^2} \rho_k(u_k, v_k) \log \rho_k(u_k,v_k)\,
du_k
dv_k  \\
& = & -\sum_{k=1}^n \left ( \log \frac{\betan \lambda_k}{2 \pi} -1 \right )
\\
& = & C(n) + n \log\left( \frac{L^2}{\betan} \right)\,,
\end{eqnarray*}
where $C(n) = n - \sum_{k=1}^n \log ( k^2 \pi/2)$ 
depends only on the number of Fourier modes $n$.  Here, we have used
the identity $\lambda_k = (k \pi/L)^2$.  From this 
calculation and equation (\ref{betan}), it follows that
\begin{equation} \label{entmax}
S(\rho^{(n)}) = C(n) + n \log 
\left ( \frac{ L^2\, [ H^0 - H_n ( \langle u^{(n)} \rangle,\langle v^{(n)}
\rangle )] }{n}
\right ) \,.
\end{equation}
Evidently, the entropy $ S(\rho^{(n)})$ is maximized if and only if
the mean field pair $(\langle u^{(n)} \rangle, \langle v^{(n)}
\rangle)$ corresponding to $\rho^{(n)}$ minimizes the Hamiltonian
 $H_n$ over all
fields $(u^{(n)},v^{(n)}) \in X_n \times X_n$ that satisfy the
constraint $N_n (u^{(n)} , v^{(n)}) = N^0$.  That such minimizers
exist is proved in the Appendix.

\vspace{1.5ex}

The equation (\ref{entmax}) has the interesting interpretation that, up
to additive and multiplicative constants, the entropy of the mean-field
equilibrium $\rhon$ is the logarithm of the kinetic energy
contained in the fluctuations about the mean state, which is the
coherent structure.  Because the potential energy is determined
entirely by the mean, the difference between the prescribed total
energy $H^0$ and the energy of the mean must be accounted for entirely
by the contribution of the fluctuations to the kinetic energy.  The
same conclusion also follows from equation (\ref{hameqn1}).  This
result provides a new and definite interpretation to the notion set
forth by Yankov {\em et al.} \cite{KY,DZPSY,ZPSY} and Pomeau
\cite{Pomeau} that the ``entropy'' of the NLS system is directly
related to the amount of kinetic energy contained in the small--scale
fluctuations, or radiation, and that the dynamical tendency of
solutions to approach a ground state that minimizes energy at a
conserved particle number is ``thermodynamically advantageous.''

The parameter $\lambdan$ is determined as a Lagrange multiplier for
the mean-field variational problem: minimize $H_n$ subject to the
constraint that $N_n=N^0$.  The multiplier associated with any
minimizer $ \langle \phin \rangle = \langle u^{(n)} \rangle + i
\langle v^{(n)} \rangle$ is unique, even though the minimizer itself
may be nonunique.  Indeed, the family of solutions $e^{i \theta}
\langle \phin \rangle$ for arbitrary (constant) phase shifts $\theta$
exists for every such minimizer.  Moreover, the ground state equation
is a nonlinear eigenvalue equation, whose solutions can bifurcate.
These bifurcations play the role of phase transitions in the
mean-field statistical theory.  Some analysis of the solutions of the
mean-field equation is provided in  the Appendix.

Once a solution pair $( \langle \phin \rangle ,\lambdan )$ is found,
with $H_n ( \langle \phin \rangle ) = H_n^*$ being the minimum value
of $H_n$ allowed by the particle number constraint $N_n = N^0$,
the ``inverse temperature''
$\betan$ is uniquely determined by 
\begin{equation} \label{betan2}
\beta^{(n)} = \frac{n}{H^0 - H_n^*}\,.
\end{equation}
Indeed, this is a restatement of (\ref{betan}).  Since $ H_n^*$
approaches a finite limit as $n \rightarrow \infty$, we find that
$\betan$ scales linearly with the number of modes $n$.  This
asymptotic scaling of inverse temperature, $\betan \sim n \beta^*$,
where $\beta^*$ is finite in the continuum limit, is what
distinguishes our mean-field theory from the statistical equilibrium
theories discussed in Section 3.  In those theories the inverse
temperature is fixed and the mean energy is allowed to go to infinity
in the continuum limit.  Our rescaling of inverse temperature with $n$
is necessary in order that the expectation of Hamiltonian with respect
to $\rho^{(n)}$ remain finite in the limit as $n \rightarrow \infty$.

The implications of this scaling of $\betan$ are most evident in the
the particle number and kinetic energy spectral densities.  
Substituting  the expression (\ref{betan2}) for $\betan$ into (\ref{vark}), 
we obtain the following
formulas for the variances of the Fourier components $u_k$ and $v_k$:
\begin{equation} \label{vark2}
\mbox{Var }(u_k) = \mbox{Var }(v_k) = \frac{H^0 - H_n^*}{n \lambda_k}\,.
\end{equation} 
From these equilibrium expressions we obtain  a sharp form of the
vanishing of fluctuations hypothesis (\ref{hyp1}), which we adopted to
derive the mean-field ensembles; namely,
\begin{equation} \label{partfluct}
  \frac{1}{2} \sum_{k=1}^n [ \mbox{Var}(u_k) + \mbox{Var}(v_k)]
  \;=\; \frac{H^0 - H_n^*}{n} \sum_{k=1}^n \frac{1}{\lambda_k}
  \;=\; O\left(\frac{1}{n}\right)\,\;\; \mbox{as } n \rightarrow \infty\, ,
\end{equation}
using the fact that $\lambda_k^{-1} = O (k^{-2})$ in the estimate.
Thus, we have the following prediction for the particle
number spectral density
\begin{equation} \label{partspectra} 
\frac{1}{2}\langle u_k^2 + v_k^2 \rangle = \frac{1}{2} ( \langle u_k
\rangle
^2 + \langle v_k
\rangle^2 ) + \frac{H^0 - H_n^*}{n \lambda_k}\,.
\end{equation}
Since the mean-field is a smooth solution of the ground state
equation, its spectrum decays rapidly in $k$, and we have the
approximation $\frac{1}{2}\langle u_k^2 + v_k^2 \rangle \approx (H^0 -
H_n^*)/(n \lambda_k)$ for $k >>1/L$.  In the same way, the kinetic
energy spectral density is 
\begin{equation} \label{kinspectra} 
\frac{1}{2} \lambda_k(\langle u_k^2 + v_k^2 \rangle) = 
\frac{1}{2}\lambda_k ( \langle u_k \rangle ^2 +
\langle v_k
\rangle^2 ) + \frac{H^0 - H_n^*}{n}\,,
\end{equation}
This expression shows that, as may be anticipated from a statistical
equilibrium ensemble, the contribution to the kinetic energy from the
fluctuations is equipartitioned among the $n$ spectral modes.  These
predictions are tested against the results of direct numerical
simulations in \cite{JJ}, where a good agreement with the mean-field
theory is documented.


\section{Continuum Limit of Random Fields}
\label{continuumsect}

In this section we investigate the limits as $n \rightarrow \infty$ of
the random fields $u^{(n)}$, $v^{(n)}$, and their gradients.  
The random fields $u^{(n)}$ and $v^{(n)}$ will be seen to converge in a
uniform way to deterministic limits $u^{*}$ and $v^{*}$, while the gradients
$u^{*}_{x}$ and $v^{*}_{x}$ converge in a weaker sense to random fields
with finite variance but no spatial coherence.

A simple form of the convergence of $u^{(n)}$ can be seen from the following
easy estimate:
\begin{eqnarray} \label{varn}
\mbox{ Var}(\un(x)) &=& \sum_{k=1}^n \mbox{Var}(u_k) \, (e_k(x))^2  
\nonumber \\
&=&  \frac{1}{\betan}  \sum_{k=1}^n \frac{1}{\lambda_k}
  (e_k(x))^2 \nonumber \\  &=& O(n^{-1}),
\end{eqnarray}
and similarly for $\vn(x)$.
Thus, the variance at each point goes to zero as $n \rightarrow \infty$.

A more detailed analysis is complicated by the fact that the mean fields
$(\langle u^{(n)} \rangle, \langle v^{(n)} \rangle )$ are known to converge
only on appropriately selected subsequences.  This result is the content of
Theorem A1 proved in the Appendix.  It states that such subsequences of
mean fields converge in the $C^1$-norm to ground states $(u^*, v^*)$ that
solve the continuum equation (\ref{gse}); these ground states minimize the
Hamiltonian $H$ over all $(u,v) \in H^1_0(\Omega)$ that satisfy the
particle number constraint $N(u,v) = N^0$.  The possible nonuniqueness of
solutions to the ground state equation necessitates the introduction of
subsequences, since in general there may exist a {\em set} of limits $(u^*,
v^*)$ for a specified constraint value $N^0$.  Accordingly, we shall assume
throughout this section that $n$ goes to infinity along a subsequence for
which the limiting ground state $(u^*, v^*)$ exists, without relabeling
$n$.

The detailed properties of the continuum limit are expressed in the
following theorem.

\vspace{1.5ex}
\noindent
{\bf Theorem 2.} The (sub)sequence of random fields $(u^{(n)}, v^{(n)})$
whose distribution is the mean-field ensemble with density $\rho^{(n)}$
converges in distribution to a non--random field $(u^*, v^*)$ which
minimizes the Hamiltonian $H$ given the particle number constraint $N=N^0$.
In addition, the gradients $(u^{(n)}_x, v^{(n)}_x)$ converge in the sense
of finite-dimensional distributions to a Gaussian random field $(U'(x),
V'(x))$ on $\Omega=[0,L]$ with the following properties: $U'$ and $V'$ are
statistically independent, the mean of $(U',V')$ is $(u^*_x, v^*_x)$, and
$U'$ and $V'$ each have the covariance function
\begin{equation} \label{Gcovar}
  \Gamma^*(x,y) = \frac{H^{0}-H^{*}}{L} \left\{ \begin{array}{cl}
                0, & \mbox{if $x \ne y$} \\
                1, & \mbox{if $x = y$ and $0 < x < L$} \\
                2, & \mbox{if $x = y$ and $x= 0, L$} \end{array}
                \right.
\end{equation}
where $H^*$ is the minimum value of $H$ given $N=N^0$.
\vspace{1.5ex}

{\bf Proof.} Recall that for each $n$, the mean $(\unmean, \vnmean)$
under $\rhon$ is a solution to (\ref{mfeqn}) which
minimizes $H_n$ over $X_n \times X_n$ subject to the constraint $N_n =
N^0$.  By Theorem A1, the (sub)sequence $(\unmean, \vnmean)$ converges
in $C^1(\Omega)$ to $(u^{*}, v^{*})$, where $(u^{*}, v^{*})$ is a
solution of (\ref{gse2}) that minimizes $H$ subject to the constraint
$N(u,v) = N^0$. 

Recalling (\ref{vark}) and the independence of the Fourier coefficients
$u_{k}$, we may use the following representation of $u^{(n)}$:
\begin{eqnarray} \label{gg}
    u^{(n)}
     =  \sum_{k=1}^{n} u_{k} e_{k}
    & = & \sum_{k=1}^{n} \left [ \langle u_{k} \rangle +
          \frac{1}{\sqrt{\lambda_{k} \betan}} Z_{k} \right ]
 e_{k} \nonumber \\
    & = & \langle u^{(n)} \rangle + \frac{1}{\sqrt{\betan}} \sum_{k=1}^{n}
          \frac{L}{\pi k} Z_{k} e_{k},
\end{eqnarray}
where $Z_{1}, Z_{2}, \ldots$ are independent Gaussian random variables
with mean 0 and variance 1.
The sum in the last line is the $n^{\mbox{th}}$ partial sum of the
Karhunen-Lo\`{e}ve expansion of the Brownian bridge on $\Omega$, which is
known to converge to its limit $B(x), x \in \Omega$, uniformly with
probability one as $n \rightarrow \infty$ (\cite{Adler}, Theorem 3.3.3).
That is, for each  function $h: C(\Omega) \rightarrow R$ which
is continuous in the supremum norm,
 we have, with probability one,
\begin{equation}
    h\left(\sum_{k=1}^{n} \frac{L}{\pi k} Z_{k} e_{k}\right) 
    \rightarrow h(B) \;\; \mbox{as} \;\; n \rightarrow \infty.
\end{equation}
It is clear from this demonstration that for such $h$ we have
\begin{equation}
    h(u^{(n)}) \rightarrow h(u^{*}) \;\; \mbox{as} \;\; n \rightarrow \infty
\end{equation}
with probability one, since along this (sub)sequence the means converge
uniformly and $\beta^{(n)}$ goes to infinity.
Restricting to bounded $h$ and taking averages, we have
\begin{equation}
   \langle h(u^{(n)}) \rangle \rightarrow h(u^{*})
    \;\; \mbox{as} \;\; n \rightarrow \infty,
\end{equation}
which is the definition of convergence in distribution of the random fields
$u^{(n)}$ to $u^{*}$.
The argument that $v^{(n)}$ converges to $v^{*}$ is identical.

The analysis of the gradient fields $u^{(n)}_{x}$ and $v^{(n)}_{x}$ is
simpler.
First note that they are independent and differ only in their means, and
consequently it is sufficient to consider $u^{(n)}_{x}$, say.  From 
the expression
\begin{equation}
    u^{(n)}_{x}(x) = \sum_{k=1}^{n} u_{k} (e_{k})_{x}(x) \, ,
\end{equation}
it is evident that $u^{(n)}_{x}$ is a continuous Gaussian random field
whose mean $\ungrad$ converges uniformly to $u^{*}_{x}$ as $n \rightarrow
\infty$.
Its covariance is then calculated to be
\begin{eqnarray} \label{covargradn1}
 \Gamma_n(x,y)
 & = &  \frac{1}{\betan} \sum_{k=1}^n \frac{1}{\lambda_k}
        (e_k)_{x}(x)(e_k)_{x}(y)\nonumber \\
 & = & \frac{1}{2 L \betan } \left [ D_n ( \pi (x+y)/L)
        + D_n ( \pi(x-y)/L ) -2 \right ]\,,
\end{eqnarray}
where
\begin{displaymath}
   D_n(\theta) =
   \left\{
   \begin{array}{cl}
   \frac{\sin (n + \frac{1}{2})\theta}{\sin\frac{1}{2}\theta}, &
      \mbox{for } \theta \neq 0, \pm 2\pi, \pm 4\pi, \ldots \\
   2n +1, & \mbox{for } \theta = 0, \pm 2\pi, \pm 4\pi, \ldots
   \end{array} \right.
\end{displaymath}
is the Dirichlet kernel \cite{CHQZ}. The asymptotic behavior
$\betan \sim n/(H^0 - H^*)$ implied by (\ref{betan2}) combined with
the properties of
$D_n(\theta)$ then produce the desired result that,
as $n \rightarrow \infty$, the covariance function
$\Gamma_n(x,y)$ converges pointwise to the function $\Gamma^*(x,y)$
given in the statement of the theorem.

Now let $U'$ be the Gaussian random field with mean $u^{*}_{x}$ and
covariance $\Gamma^{*}$.
For convergence of finite-dimensional distributions, it remains to show
that for each $m$ and any $x_{1}, \ldots, x_{m}$ in $\Omega$, the joint
distribution on $R^{m}$ of the vector $(u^{(n)}_{x}(x_{1}), \ldots,
u^{(n)}_{x}(x_{m}))$ converges weakly to the joint distribution of
$(U'(x_{1}), \ldots, U'(x_{m}))$ as $n \rightarrow \infty$.
But this follows by the L\`{e}vy Continuity Theorem \cite{Billingsley},
since the characteristic function of the former vector converges to the
characteristic function of the latter vector, simply because the means and
covariances converge.
The argument that $v^{(n)}_{x}$ converges to $V'$ is similar, as is the
extension to the convergence of the pair $(u^{(n)}_{x}, v^{(n)}_{x})$.

\vspace{1.5ex}

The results of Theorem 2 may be paraphrased as follows:
The mean-field distributions
$\rhon$ converge weakly as measures on $C(\Omega)$ (along a subsequence) to a
Dirac mass concentrated at a minimizer of the Hamiltonian subject to the
particle number constraint.
In this limit, the variance of the field $\phin$ vanishes while the
variance of its gradient $\phin_x$ is uniform over the interval $\Omega$,
apart from the endpoints.
This behavior is in good qualitative agreement with the long--time behavior
observed in numerical simulations \cite{JJ}.

Nonetheless, the continuum limit is degenerate in two ways.  First, the
limiting field $\phi^*$ is deterministic.
Second, the limiting gradient field, whose mean is $\phi^*_x$ and
covariance function is $\Gamma^*$, is so rough that it does not even take
values in the space of measurable functions (that is, it does not have a
measurable version) \cite{RevuzYor}.
These unusual properties of the limiting distribution are a direct
consequence of the scaling of the inverse temperature $\betan$ with $n$,
which is required to maintain finite mean energy in the continuum limit.

\section{The Concentration Property}
\label{concpropsect}

We shall establish in this section the important result that the
mean-field ensembles $\rhon$, which solve the maximum entropy principle
(MEP), become equivalent in a certain sense to the microcanonical
ensemble (\ref{microcan})
in the continuum limit $n \rightarrow \infty$. Specifically, we shall
prove the following {\em concentration property}:

\vspace{1.5ex}
\noindent
{\bf Theorem 3.} Let $\rhon, n=1,2,\cdots$, be a sequence of solutions
of the constrained variational principle (MEP).  Then

\begin{equation}\label{th3.1}
\lim_{n \rightarrow \infty} \langle N_n \rangle = N^0\,,\;\;\;
\lim_{n \rightarrow \infty} \mbox{Var}(N_n) = 0\,,
\end{equation}
and

\begin{equation}\label{th3.2}
\lim_{n \rightarrow \infty} \langle H_n \rangle = H^0\,,\;\;\;
\lim_{n \rightarrow \infty} \mbox{Var}(H_n) = 0\,.
\end{equation}

\vspace{1.5ex}

This theorem has the interpretation that the mean-field ensembles
$\rhon$ concentrate on the microcanonical constraint manifold
$N_n= N^0, H_n=H^0$ in the continuum limit.  As we have emphasized
above, it is a well-accepted axiom of statistical mechanics that the
microcanonical ensemble constitutes the appropriate statistical
equilibrium description for an isolated ergodic system
\cite{Balescu,PB}. Thus, this concentration property provides
an important theoretical justification for our mean-field theory.

In the proof of Theorem 3, we will make use of the following
elementary facts concerning Gaussian random variables:  If $W$ 
is a Gaussian random variable with mean $\mu$ and variance
$\sigma^2$, then
\begin{equation} \label{gaussfact}
\langle ( W - \mu )^4 \rangle = 3 \sigma^4\,,\;\;\;
\langle W^4 \rangle - \langle W^2 \rangle^2 =
2 \sigma^4 + 4 \sigma^2 \mu^2\,.
\end{equation}  
We will also repeatedly make use of the result 
that $1/\betan = O(n^{-1})$, which follows from equation (\ref{betan2}).

\vspace{1.5ex}
\noindent
{\bf Proof of Theorem 3: } The first conclusion in 
(\ref{th3.1}) is easy to
establish. Indeed, using  (\ref{mfc1}) and the calculation
(\ref{partfluct}), we have 
\begin{eqnarray*}
\langle N_n \rangle & = &\frac{1}{2} \sum_{k=1}^n (\langle u_k \rangle^2
+ \langle v_k \rangle^2 ) + 
\sum_{k=1}^n [ \mbox{Var}(u_k) + \mbox{Var}(v_k) ] \\
& = & N^0 + O(n^{-1})\,,\; \mbox{as } n \rightarrow \infty\,.
\end{eqnarray*}
To verify the second assertion in (\ref{th3.1}), we calculate using
(\ref{dispn})

\begin{eqnarray} \label{th3.3}
\mbox{Var}(N_n) & = & \frac{1}{4} \left \langle \left (
\sum_{k=1}^n \left [ ( u_k^2 - \langle u_k^2 \rangle )
+ ( v_k^2 - \langle v_k^2 \rangle ) \right ] \right )^2 \right \rangle
 \nonumber \\
& = & \frac{1}{4} \sum_{k=1}^n 
\left [ (\langle u_k^4 \rangle - \langle u_k^2 \rangle^2 ) +
(\langle v_k^4 \rangle - \langle v_k^2 \rangle^2) \right ]\,,
\end{eqnarray}
where, to obtain the second equality in this calculation, we have
made use of the mutual statistical independence of the random
variables $u_1,\ldots,u_n, v_1, \ldots, v_n$. But, as $u_k$ and $v_k$
are Gaussian with means $\langle u_k \rangle$
and $\langle v_k \rangle$ and variances given by (\ref{vark2}), we
have from (\ref{gaussfact}) that

\begin{equation} \label{th3.4}
\langle u_k^4 \rangle - \langle u_k^2 \rangle^2 = \frac{2}{(\betan
\lambda_k)^2} + 
\frac{4 \langle u_k \rangle^2}{\betan \lambda_k}\,,\;\;
\langle v_k^4 \rangle - \langle v_k^2 \rangle^2 = \frac{2}{(\betan
\lambda_k)^2} + 
\frac{4 \langle v_k \rangle^2}{\betan \lambda_k}\,.
\end{equation}
Upon substituting (\ref{th3.4}) into (\ref{th3.3}), we obtain

\begin{eqnarray*}
\mbox{Var}(N_n) & =& \frac{1}{(\betan)^2} \sum_{k=1}^n
\frac{1}{\lambda_k^2} +
\frac{1}{\betan}\sum_{k=1}^n
 \left [ \frac {\langle u_k \rangle^2 +\langle v_k \rangle^2}
{\lambda_k} \right ] \\
& \rightarrow & 0\;\;\; \mbox{as }\; n \rightarrow \infty\,,
\end{eqnarray*}
since $\sum_{k=1}^n \lambda_k^{-2}$ converges as $n \rightarrow
\infty$, and $\sum_{k=1}^n
\lambda_k^{-1} ( \langle u_k \rangle^2 +\langle v_k \rangle^2 ) \le
2\lambda_1^{-1} N^0$ for all $n$.
  
To establish the first part of (\ref{th3.2}), observe that
 from (\ref{mfc2}), there holds 

\begin{equation}\label{th3.4.5}
\left | \langle H_n \rangle - H^0 \right | = 
\left | \langle \Theta_n(\un, \vn) \rangle
- \Theta_n (\langle \un \rangle, \langle \vn \rangle) \right |\,,
\end{equation}
where the potential energy $\Theta_n$ is defined by (\ref{tn}).
Mimicking the calculations following equation (\ref{Napprox})
and preceding equation (\ref{Happrox}), we find that

\begin{equation}\label{th3.5}
\left | \langle \Theta_n(\un, \vn) \rangle
- \Theta_n (\langle \un \rangle, \langle \vn \rangle) \right |
  \le C \sum_{k=1}^n \left [ 
\mbox{Var}(u_k) + \mbox{Var}(v_k) \right ]\,,
\end{equation}
where $C$ is a constant independent of $n$. Thus, according to
(\ref{th3.4.5}), (\ref{th3.5}) and (\ref{partfluct}), there holds

\begin{displaymath}
\left | \langle H_n \rangle - H^0 \right | = O(n^{-1})\,,\;\;
\mbox{as }\, n \rightarrow \infty\,.
\end{displaymath}

Next, we have that

\begin{equation} \label{th3.7}
\mbox{Var}(H_n) = \mbox{Var}( K_n + \Theta_n) \le 2\; 
\mbox{Var}(K_n) + 2\; \mbox{Var}(\Theta_n)\,,
\end{equation}
where $K_n$ is the kinetic energy defined as in (\ref{kn}).
Using the statistical independence properties noted above, we calculate

\begin{eqnarray*}
\mbox{Var}(K_n) & = &
\frac{1}{4} \left \langle \left ( \sum_{k=1}^n \lambda_k \left [ (u_k^2 -
\langle u_k^2 \rangle)
+ (v_k^2 - \langle v_k^2 \rangle) \right ] \right )^2 \right \rangle \\  
& = &
\frac{1}{4} \sum_{k=1}^n \lambda_k^2 \left [ (\langle u_k^4 \rangle - \langle
u_k^2 \rangle^2)
+ (\langle v_k^4 \rangle - \langle v_k^2 \rangle^2) \right ]\,,
\end{eqnarray*}
and using (\ref{th3.4}), we arrive at

\begin{displaymath} 
\mbox{Var}(K_n) =\frac{n}{(\betan)^2} + \frac{1}{\betan} 
\sum_{k=1}^n \lambda_k  (\langle u_k \rangle^2 +
\langle v_k \rangle^2)\,.
\end{displaymath}
This last expression vanishes as $n \rightarrow \infty$,
 because the sum $\sum_{k=1}^n \lambda_k (\langle u_k \rangle^2 +
\langle v_k \rangle^2)$, which represents twice the kinetic energy of
the mean, is bounded independently of $n$.

Finally, we show that the variance of the potential energy $\Theta_n$ 
tends to 0 in the continuum limit. Using the definition of $\Theta_n$
and the Cauchy--Schwarz inequality, we have

\begin{eqnarray} \label{th3.8}
\mbox{Var}(\Theta_n) & = & \frac{1}{4}\left \langle \left (
\int_{\Omega} \left [ F((\un)^2 + (\vn)^2) - 
\left \langle  F((\un)^2 + (\vn)^2)
\right \rangle  \right ]\, dx \right )^2 \right \rangle \nonumber \\
& \le & \frac{L}{4} \int_{\Omega} 
\mbox{Var}\left ( F((\un)^2 + (\vn)^2) \right ) \, dx\,.
\end{eqnarray}
Now the potential $F$ may be expanded about the mean
 $(\langle \un \rangle, \langle \vn \rangle)$ as 

\begin{eqnarray*}
F((\un)^2 + (\vn)^2)& = & F(\langle \un \rangle^2 + \langle
\vn\rangle^2)
\\
& + & 2 f(\langle u^{(n)} \rangle^2 + \langle v^{(n)}
  \rangle^2)\left [ \langle u^{(n)} \rangle (u^{(n)} - \langle u^{(n)}
    \rangle ) + \langle v^{(n)} \rangle (v^{(n)} - \langle v^{(n)}
    \rangle ) \right ] \\
&+& \frac{1}{2}
  \left ( \begin{array}{c} (u^{(n)} - \langle u^{(n)}
      \rangle) \\ (v^{(n)} - \langle v^{(n)} \rangle) \end{array}
  \right )^{T} J(\tilde{u}^{(n)}, \tilde{v}^{(n)}) \left (
    \begin{array}{c} (u^{(n)} - \langle u^{(n)} \rangle) \\ (v^{(n)} -
      \langle v^{(n)} \rangle) \end{array} \right )\,,
\end{eqnarray*}
where the matrix $J$ is defined
by (\ref{matrix}) and
 $(\tilde{u}^{(n)}, \tilde{v}^{(n)})$ lies between $(\un, \vn)$
and $(\langle \un \rangle, \langle \vn \rangle)$. Using the
boundedness condition (\ref{boundcond}) on $f$ and the statistical 
independence of $\un$ and $\vn$, we obtain after a straightforward
analysis, that pointwise over $\Omega$ 

\begin{eqnarray} \label{th3.9}
\mbox{Var}\left ( F((\un)^2 + (\vn)^2) \right )
&\le & C \left [  \langle u^{(n)} \rangle^2\; \mbox{Var}(\un)
+  \langle v^{(n)} \rangle^2\; \mbox{Var}(\vn) \right ] \nonumber \\
& + & C \left [\langle ( u^{(n)} - \langle u^{(n)} \rangle
)^4 \rangle
+ \langle ( v^{(n)} - \langle v^{(n)} \rangle )^4 \rangle \right ]\,.
\end{eqnarray}
Here, $C$ is a constant that is independent of both $n$ and $x \in \Omega$.
As we have
demonstrated in (\ref{varn}), 
\begin{equation} \label{th3.10}
\mbox{Var}(\un (x)) = \mbox{Var}(\vn(x)) =
\frac{1}{\betan} \sum_{k=1}^n \frac{e_k^2(x)}{\lambda_k}\,,
\end{equation}
 and since  $\un(x)$ and $\vn(x)$ are Gaussian variables, it follows from
(\ref{th3.10})
and
(\ref{gaussfact}) that 

\begin{equation}\label{th3.11}
\left \langle ( u^{(n)}(x) - \langle u^{(n)}(x) \rangle )^4 \right \rangle =
\left \langle ( v^{(n)}(x) - \langle v^{(n)}(x) \rangle )^4 \right \rangle  
= \frac{3}{(\betan)^2} \left ( \sum_{k=1}^n \frac{e_k^2(x)}{\lambda_k}
\right )^2\,.
\end{equation} 
Thus, from (\ref{th3.8})--(\ref{th3.11}), the fact that
the eigenfunctions $e_k$  are uniformly bounded over
$\Omega$ independently of $k$, and the fact that  
$\int_{\Omega} ( \langle u^{(n)} \rangle^2 + 
\langle v^{(n)} \rangle^2 )\, dx = 2N^0$ for all $n$, 
 we have 
\begin{displaymath}
\mbox{Var}(\Theta_n) = O(n^{-1})\,,\;\; \mbox{as }\; 
n \rightarrow \infty.
\end{displaymath}

This completes the proof of Theorem 3.


\section{Extension to Unbounded Nonlinearities}
\label{unboundedsect}

\vspace{1.5ex}  

In this section, we briefly indicate how our mean-field statistical 
 theory,  can be extended  to a class
of unbounded nonlinearities, which includes the focusing
power law nonlinearities $f(|\psi|^2) = |\psi|^s, 0 < s < 4$.
As illustrated in \cite{JJ,DZPSY,ZPSY},
numerical simulations for such power law nonlinearities
exhibit the same phenomena as seen for bounded nonlinearities such as
$f(|\psi|^2) = |\psi|^2/(1+|\psi|^2)$, for which the results of numerical
simulations
are displayed in Figures (1)-(2). That is,
 the field $\psi$ approaches a long-time state consisting of
a coherent soliton structure coupled with 
radiation or fluctuations of very small
amplitude. As in the case of the bounded nonlinearities, 
the gradient $\psi_x$ exhibits
fluctuations of nonnegligible amplitude for all time.  In fact, we expect,
that this general behavior occurs as long as the nonlinearity $f$ is such
that
the NLS equation 
(\ref{nls1}) is nonintegrable and free of collapse 
(i.e., such that finite time singularity
does not occur).  Hence, we would like to 
apply our statistical theory to  NLS with such nonlinearities, as well. 

Let us recall that to motivate the mean-field constraints (\ref{mfc1})
and (\ref{mfc2}) for nonlinearities $f$ satisfying the boundedness
condition ({\ref{boundcond}), we invoked the vanishing of 
fluctuations hypothesis (\ref{hyp1}).  However, for nonlinearities
such that the potential $F(|\psi|^2)$ grows more rapidly than
$C |\psi|^2$ as $|\psi| \rightarrow \infty$ (as it does for the power
law nonlinearities), the hypothesis
(\ref{hyp1}) is not sufficient to guarantee {\em  a priori} that $\langle
\Theta_n (\un,\vn) \rangle$ converges to 
$\Theta_n (\langle \un \rangle, \langle \vn \rangle)$ as $n
\rightarrow \infty$.  Thus the mean-field Hamiltonian constraint
(\ref{mfc2}) can
not be
derived from the vanishing of fluctuations hypothesis (\ref{hyp1})
alone for such $f$.  Note that by making a stronger vanishing of
fluctuations hypothesis, we could have weakened the assumptions on $f$
and arrived at the same mean-field constraints (\ref{mfc1})-(\ref{mfc2}). 
In any case, we could simply impose these mean-field constraints and
investigate the resulting maximum entropy ensembles $\rhon$.  If it
can be shown that these ensembles exist and satisfy the concentration 
property expressed in Theorem 3, then we will consider this approach
to be justified {\em a posteriori}.
 
It is clear from Theorem 1 and Theorem 2 that, in order for our
statistical theory to be well-defined, it is necessary that there
exist minimizers (in $H^1_0(\Omega)$, say) 
of the Hamiltonian $H$ given the particle number
constraint $N=N^0$.  This places
restrictions on the class of nonlinearities that we can consider.
However for nonlinearities $f$ such that the potential $F$ satisfies
\begin{equation} \label{potcond}
 \mbox{There exists } C > 0\;\; \mbox{such that }
F(|\psi|^2) \le C (|\psi|^2 + |\psi|^q )\,,\; \mbox{for some }\; 
2 \le q < 6\,,
\end{equation}
it may be shown that such minimizers exist for any $N^0$
\cite{CL}.  The condition (\ref{potcond}) is also crucial in
establishing the well-posedness of the NLS equation (\ref{nls1})
as an initial value problem in the Sobolev space $H^1_0(\Omega)$
\cite{strauss}. Clearly, the focusing power law nonlinearities 
$f(|\psi|^2) = |\psi|^s$ satisfy this condition when $0 < s < 4$.

Thus, assuming that $F$ satisfies the growth condition
(\ref{potcond}),
we may investigate the ensembles $\rhon$ which maximize entropy
subject to the mean-field constraints (\ref{mfc1})-(\ref{mfc2}).
It is easy to see that the analysis 
of Section 6 goes through without
change. Thus, under the the maximum entropy ensemble $\rhon$, the
Fourier coefficients $u_1, \cdots\, u_n, v_1, \cdots\, v_n$ are
mutually
independent Gaussian variables,  the mean-field minimizes the
Hamiltonian given the particle number constraint $N_n = N^0$, and the
variances of the $u_k$ and $v_k$ are given by (\ref{vark2}).
It may be shown that the continuum limit of Section 7 and the 
important concentration property of Section 8 also hold, if we impose,
in addition to (\ref{potcond}), a lower bound on the potential $F$ of
the form
\begin{equation} \label{potcond2}
\mbox{There exists } c \ge 0\;\; \mbox{such that } 
-c |\psi|^r \le F(|\psi|^2)\,,\; \mbox{for some } r \ge 0\,.
\end{equation}
That is, Theorems 2 and 3 still hold for $F$ satisfying 
(\ref{potcond}) and (\ref{potcond2}), but the proofs are complicated
by the unboundedness of $f$.  Since our goal in this paper
has been to emphasize conceptual issues rather than technical details,
we shall not present the proofs here.  However, we refer the
interested reader to \cite{jtz2}, where some of the analysis has been
carried out for nonlinearities satisfying (\ref{potcond}) and
(\ref{potcond2}).

\section*{Acknowledgments} 

The research of R. Jordan has been
supported in part by an NSF Postdoctoral
Research Fellowship and also by DOE through
grants to the Center for Nonlinear Studies at 
Los Alamos National Laboratory.  The research of B. Turkington
is partially supported by the NSF through the grant
NSF-9600060.
R. Jordan thanks Christophe Josserand and Pieter Swart
for valuable discussions and suggestions.


\section*{Appendix}

In this Appendix, we provide an analysis of the variational principle
that defines the mean field corresponding to the maximum entropy ensemble
$\rhon$.  In particular, we prove that for (smooth)
 nonlinearities $f$ satisfying
the  condition (\ref{boundcond}),
solutions
of this variational principle exist,
and we investigate the convergence properties
of these
mean fields as $n \rightarrow \infty$.  

First, we consider the variational principle
\begin{equation} \label{VP}
H(\psi) \rightarrow \mbox{min}\,,\;\;
\mbox{subject to } \psi \in A^0\,,
\end{equation}
where the admissible set $A^0$ is defined as
\begin{equation} \label{A0}
A^0 = \left \{ \psi :\Omega \rightarrow {\bf C}\; |\; \psi \in
H^1_0(\Omega)\,,\; N(\psi) = N^0
\right \}\,.
\end{equation}
The Hamiltonian $H$ and the particle number $N$ are defined by equations
(\ref{ham1}) and (\ref{part1}), respectively, and, as in the main text,
$\Omega=[0,L]$.  Recall also that the potential $F$ is defined
as $F(a) = \int_0^a f(a')\, da'$.

\vspace{1.5ex}
\noindent
{\bf Lemma A1.} There exists
a solution $\psi \in H^1_0(\Omega)$ of the variational problem (\ref{VP}).

\vspace{1.5ex}

\noindent
{\bf Proof:} As $f$ satisfies the boundedness condition
(\ref{boundcond}),
there exists a constant $K > 0$ such that $|F(|\psi|^2)| \le K
|\psi|^2$. Thus, for all $\psi \in A^0$, there holds
\begin{eqnarray} \label{a1}
H(\psi) & = &  \frac{1}{2} \int |\psi_x|^2\, dx - 
\frac{1}{2}\int F(|\psi|^2)\, dx  \nonumber \\
& \ge & \frac{1}{2} \int |\psi_x|^2\, dx - \frac{K}{2}N^0\,.
\end{eqnarray}
Here, and in the remainder of
 this Appendix, all integrals are over the domain $\Omega$.
It follows from (\ref{a1}) that $H$ is bounded below on $A^0$, so that we may  
choose
a minimizing sequence $ (\psi_n)_{n \in {\bf N}} \subset A^0$ for $H$.
The inequality (\ref{a1}) also implies that the sequence $\psi_n$ is
uniformly bounded in $H^1_0(\Omega)$. 
 Hence, by the Sobolev Imbedding Theorem \cite{Adams},
there exists a subsequence, still denoted by $\psi_n$,  
and a function $\psi$ such that
$\psi_n$ converges weakly to $\psi$ in $H^1_0(\Omega)$ and $\psi_n$
converges
in $C(\Omega)$ to $\psi$.  Because of the convergence in $C(\Omega)$,
we have that
$\int F(|\psi|^2)\,dx = \lim_{n \rightarrow \infty} \int
F(|\psi_n|^2)\,dx$.
Also, since $\psi_n$ converges weakly to $\psi$ in $H^1_0(\Omega)$,
there holds
$\int |\psi_x|^2\, dx \le \liminf_{n \rightarrow \infty} \int
|(\psi_n)_x|^2\,dx$.
Putting these  two results together, we obtain that 
\begin{equation} \label{a2}
H(\psi) \le \liminf_{n \rightarrow \infty} H(\psi^{(n)}) =
\inf_{\phi \in A^0} H(\phi)\,.
\end{equation}
 But, as $\psi_n$ converges to $\psi$ in $C(\Omega)$, it follows that
$N(\psi)=N^0$, so that $\psi \in A^0$.  We conclude from (\ref{a2}), 
therefore, that
$\psi$ is a solution of the variational problem (\ref{VP}).

\vspace{1.5ex}
\noindent
{\bf Lemma A2.} Any solution $\psi$ of the variational problem (\ref{VP}) is a
solution of the
ground state equation (\ref{gse}) for some real $\lambda$.

\vspace{1.5ex}

\noindent
{\bf Proof: } The conclusion of this lemma follows from an application of
the
Lagrange
multiplier rule for constrained optimization.  
The parameter $\lambda$ in the ground state equation (\ref{gse}) is the 
Lagrange multiplier which enforces the constraint $N(\psi)=N^0$.

\vspace{1.5ex}

We have demonstrated in Section (\ref{approxsect}) that
the mean fields $\langle \phi^{(n)} \rangle = \langle u^{(n)} \rangle
+ i \langle v^{(n)} \rangle $ corresponding to the maximum entropy
ensembles
$\rhon$ are solutions of the following variational problem 

\begin{equation} \label{VPn}
H(\psi) \rightarrow \mbox{min}\,,\;\;
\mbox{subject to } \psi \in A_n^0\,,
\end{equation}
where the admissible set $A_n^0$ is defined as
\begin{equation} \label{An0}
A_n^0 = \left \{ \psi \in \Xi_n\; |\; N(\psi) = N^0 \right \}\,.
\end{equation}
Here, $\Xi_n$ is the set of all complex fields $\psi$ on
$\Omega$ having the form $\psi = u + iv$, where the real fields
$u$ and $v$ are elements of the $n$-dimensional space $X_n$ spanned 
by the first $n$ eigenfunctions $e_1, \cdots, e_n$ of the operator
$-d^2/dx^2$
on $\Omega$ with homogeneous Dirichlet boundary conditions.
$\Xi_n$ may be thought of as a closed subspace of the Sobolev space
$H^1_0(\Omega)$. 
The sequence of variational problems (\ref{VPn}) corresponds to a
Ritz--Galerkin scheme for approximating solutions of the variational
principle
(\ref{VP}). The proof of the next
 lemma follows from arguments analogous to those
that were used to prove Lemmas A1 and A2.  The conclusion about the
smoothness
of solutions of (\ref{VPn}) is obvious, because any  such solution
is necessarily  a finite linear combination of the basis functions
$e_k(x)=\sqrt{2/L} \sin(k \pi x/L)$.

\vspace{1.5ex}
\noindent
{\bf Lemma A3.} For each
$n$, there exist solutions of the variational problem (\ref{VPn}).  Any
such solution
$\psi^{(n)}$ is a  solution of the differential equation
\begin{equation} \label{gsen}
 \psi^{(n)}_{xx} + P^{(n)} \left (f(|\psi^{(n)}|^2)
 \psi^{(n)} \right ) + \lambda^{(n)} \psi^{(n)} = 0\,,
\end{equation}
where $P^{(n)}$ is the projection from $H^1_0(\Omega)$ onto $\Xi_n$, and
$\lambda^{(n)}$ is
a Lagrange multiplier which enforces the constraint $N(\psi^{(n)})=N^0$. In
addition,
any solution $\psi^{(n)}$ of (\ref{VPn}) is in $C^{\infty}(\Omega)$.

\vspace{1.5ex}

We wish to investigate the convergence of solutions of (\ref{VPn}) to
those of
(\ref{VP}).  It is not our objective to obtain the strongest possible
convergence results.
For our purposes, it is sufficient to prove that any sequence of solutions
of
(\ref{VPn}) has a subsequence which converges in $C^1(\Omega)$ to a
solution
of
(\ref{VP}). Such a result is all that is needed for the analysis of the
continuum
limit of the mean field ensembles $\rho^{(n)}$ in Section
(\ref{continuumsect}). In general, we do not know that solutions of
the variational principle are unique, 
so that subsequences can not be avoided.
We shall now prove the following theorem.

\vspace{1.5ex}
\noindent
{\bf Theorem A1.} If
$\psi^{(n)}, n=1,2,\cdots$ is a sequence of solutions of the variational
problem
(\ref{VPn}),  then there exists a subsequence $\psi^{(n')}$ 
of $\psi^{(n)}$  which converges in $C^1({\Omega})$  as $n' \rightarrow
\infty$
to a solution of the variational
problem (\ref{VP}).

\vspace{1.5ex}
\noindent
{\bf Proof:} Let $\psin$ be a sequence of solutions of (\ref{VPn}), and
let $H_n^* = H(\psin)$.  Because $A_n^0 \subset A_{n+1}^0, n=1,2,\cdots$,
and
$A^0 = {\overline{\bigcup_{n=1}^\infty A_n^0}}$, we have that $H_n^* \ge
H_{n+1}^*$,
and $H_n^* \searrow H^*$ as $n \rightarrow \infty$, where $H^*$ is the
minimum value
of $H$ over $A^0$. Thus, $\psin$ is a minimizing sequence for the
variational
problem (\ref{VP}), and so following the proof of Lemma A1, there exists a
subsequence
$\psi^{(n')}$ and a solution $\psi$ of the variational
problem (\ref{VP}) such that $\psi^{(n')}$ converges weakly in
$H^1_0(\Omega)$
to $\psi$ and $\psi^{(n')}$ converges in $C(\Omega)$ 
to $\psi$. 

Now, as $\psi^{(n')}$ is a solution of the variational problem
(\ref{VPn}), it satisfies equation (\ref{gsen}) for some Lagrange
multiplier
$\lambda^{(n')}$.   We shall now show that the sequence $\lambda^{(n')},
n'
\rightarrow \infty$ is 
bounded independently of $n'$. For this purpose, we multiply equation
(\ref{gsen})
for $\psi^{(n')}$ by the  complex conjugate ${\overline{\psi^{(n')}}}$ and
integrate over
$\Omega$ to obtain
\begin{equation} \label{thA11}
-\int |(\psi^{(n')})_x|^2\, dx + \int P^{(n')} ( f(|\psi^{(n')}|^2)
\psi^{(n')})
{\overline{\psi^{(n')}}}\, dx + \lambda^{(n')} \int |\psi^{(n')}|^2\, dx =
0\,.
\end{equation}
It follows immediately from (\ref{thA11}) and the fact that $\int
|\psi^{(n')}|^2\, dx =
2N^0$ that
\begin{equation} \label{thA12}
|\lambda^{(n')}| \le \frac{1}{2N^0} \left [ \int |(\psi^{(n')})_x|^2\, dx
+
\left |\int P^{(n')} ( f(|\psi^{(n')}|^2) \psi^{(n')})
{\overline{\psi^{(n')}}}\, dx \right | \right ]\,. 
\end{equation}
Because $\psi^{(n')}$ converges
weakly to $\psi$ in $H^1_0(\Omega)$, the first integral
on the right hand side of (\ref{thA12}) is bounded independently
of $n'$.  The second integral on
the right hand side of (\ref{thA12}) may be estimated as follows: 
\begin{eqnarray*}
\left |\int P^{(n')} ( f(|\psi^{(n')}|^2) \psi^{(n')})
{\overline{\psi^{(n')}}}\, dx \right |  &\le &
\left ( \int |\psi^{(n')}|^2\, dx \right )^{\frac{1}{2}}
\left ( \int |P^{(n')} ( f(|\psi^{(n')}|^2) \psi^{(n')})|^2\, dx
\right)^{\frac{1}{2}} \\
& \le & 
\left ( \int |\psi^{(n')}|^2\, dx \right )^{\frac{1}{2}}
\left ( \int (f(|\psi^{(n')}|^2))^2 |\psi^{(n')})|^2\, dx
\right)^{\frac{1}{2}} \\
&\le & 2N^0  \sup_{x \in \Omega} |f(|\psi^{(n')}|^2)|\,.
\end{eqnarray*}
 To obtain the first line of this 
display, we have used the Cauchy--Schwarz inequality. The second line
follows from the fact that $\int |P^{(n)} \phi|^2\, dx \le \int |\phi|^2\,
dx$
for all $\phi \in H^1_0(\Omega), n \in {\bf N}$, 
and to obtain the third line, we have once again used the identity
$\int |\psi^{(n')}|^2\, dx =2N^0$.  Now, owing to
(\ref{boundcond}), 
$ \sup_{x \in \Omega} |f(|\psi^{(n')}|^2)|$ is bounded independently of
$n'$,
so the  preceding calculation demonstrates
that the second integral on the right hand side of (\ref{thA12}) can also
be
bounded independently of $n'$.  This implies that $\lambda^{(n')}$ is
uniformly
bounded in $n'$. 

The strategy now is to use the uniform boundedness
of the eigenvalues $\lambda^{(n')}$ to establish that the subsequence
$\psi^{(n')}$ is uniformly bounded in the Sobolev space $H^2_0(\Omega)$.
From
this result and the Sobolev Imbedding Theorem \cite{Adams}, we may
conclude
that the $\psi^{(n')}$ converges in $C^1(\Omega)$ to the function $\psi$
as
above, which
is a solution of the variational problem (\ref{VP}). This is the desired
conclusion.

To prove that $\psi^{(n')}$ is, in fact, uniformly bounded in
$H^2_0(\Omega)$,
we multiply the equation (\ref{gsen}) for $\psi^{(n')}$ by the
complex conjugate of $(\psi^{(n')})_{xx}$ and integrate over $\Omega$.
This yields, after integrating by parts and using the homogeneous 
Dirichlet boundary conditions,

\begin{equation} \label{thA13}
\int |(\psi^{(n')})_{xx}|^2\, dx = \lambda^{(n')} \int
|(\psi^{(n')})_x|^2\,
dx
+ \int {\overline{\psi^{(n')}}}_x  (P^{(n')} ( f(|\psi^{(n')}|^2)
\psi^{(n')}))_x\, dx\,.
\end{equation}
 Using the Cauchy--Schwarz inequality, and the
inequality $\int |(P^{n}(\phi))_x|^2\, dx \le \int |\phi_x|^2\, dx$, which
is
valid for
all $ n\in {\bf N}$ and all
$\phi \in H^1_0(\Omega)$, we find that
\begin{equation} \label{thA14}
\left |\int {\overline{\psi^{(n')}}}_x  (P^{(n')} ( f(|\psi^{(n')}|^2)
\psi^{(n')}))_x\, dx \right |
\le
\left (\int |(\psi^{(n')})_x|^2\, dx \right )^{\frac{1}{2}} 
\left ( \int  |(f(|\psi^{(n')}|^2) \psi^{(n')})_x|^2\, dx \right
)^{\frac{1}{2}}\,.
\end{equation}
But, a straightforward calculation yields that, pointwise on $\Omega$,
\begin{eqnarray} \label{thA15}
 |(f(|\psi^{(n')}|^2) \psi^{(n')})_x|^2 &\le& 2 (f(|\psi^{(n')}|^2))^2
|(\psi^{(n')})_x|^2
+8 (f'(|\psi^{(n')}|^2))^2|\psi^{(n')}|^4 |(\psi^{(n')})_x|^2 \nonumber \\
&\le& C |(\psi^{(n')})_x|^2\,,
\end{eqnarray}
for a constant $C$ independent of $n'$.   The second line of this
calculation follows from the boundedness condition (\ref{boundcond})
on $f$. Taken together, (\ref{thA13})-(\ref{thA15}) imply that
\begin{equation}\label{thA16}
\int |(\psi^{(n')})_{xx}|^2\, dx \le (C + |\lambda^{(n')}|) \int
|(\psi^{(n')})_x|^2\, dx\,.
\end{equation}
 As we have established above, $\lambda^{(n')}$ and $\int
|(\psi^{(n')})_x|^2\, dx$ are bounded
independently of $n'$, so that from (\ref{thA16}), we conclude that the
subsequence $\psi^{(n')}$
is uniformly bounded in $H^2_0(\Omega)$.  This concludes the proof of
Theorem A1.


\begin{thebibliography}{10}

\bibitem{Hasegawa}
A.~Hasegawa.
\newblock Self-organization processes in continuous media.
\newblock {\em Adv. Phys.}, 34:1, 1985.

\bibitem{McWilliams}
J.~C. McWilliams.
\newblock The emergence of isolated vortices in turbulent flow.
\newblock {\em J. Fluid Mech.}, 146:21, 1984.

\bibitem{SegreKida}
E.~Segre and S.~Kida.
\newblock Late states of incompressible 2D decaying vorticity fields.
\newblock {\em Fluid Dyn. Res.}, 23:89, 1998.

\bibitem{BW}
D.~Biskamp and H.~Welter.
\newblock Dynamics of decaying two--dimensional magnetohydrodynamic turbulence.
\newblock {\em Phys. Fluids B}, 1:1964, 1989.

\bibitem{KMT}
R.~Kinney, J.~C. McWilliams, and T.~Tajima.
\newblock Coherent structures and turbulent cascades in two--dimensional
  incompressible magnetohydrodynamic turbulence.
\newblock {\em Phys. Plasmas}, 2:3623, 1995.

\bibitem{DZPSY}
S.~Dyachenko, V.~E. Zakharov, A.~N. Pushkarev, V.~F. Shvets, and V.~V. Yan'kov.
\newblock Soliton turbulence in nonintegrable wave systems.
\newblock {\em Soviet Phys. JETP}, 69:1144, 1989.

\bibitem{ZPSY}
V.~E. Zakharov, A.~N. Pushkarev, V.~F. Shvets, and V.~V. Yan'kov.
\newblock Soliton turbulence.
\newblock {\em JETP Lett.}, 48:83, 1988.

\bibitem{JJ}
R.~Jordan and C.~Josserand.
\newblock Soliton turbulence and coherent structures
 in the nonlinear Schr\"odinger equation.
\newblock 1999.
\newblock In preparation.

\bibitem{KY}
S.~F. Krylov and V.~V. Yan'kov.
\newblock The role of solitons in strong turbulence.
\newblock {\em Soviet Phys. JETP}, 52:41, 1980.

\bibitem{Pomeau}
Y.~Pomeau.
\newblock Asymptotic time behavior of solutions of nonlinear classical field
  equations.
\newblock {\em Nonlinearity}, 5:707, 1992.

\bibitem{Balescu}
R.~Balescu.
\newblock {\em Equilibrium and Nonequilibrium Statistical Mechanics}.
\newblock Wiley, New York, 1975.

\bibitem{MWC}
J.~Miller, P.~B. Weichman, and M.~C. Cross.
\newblock Statistical mechanics, Euler equation, and Jupiter's red spot.
\newblock {\em Phys. Rev. A}, 45:2328, 1992.

\bibitem{JT}
R.~Jordan and B.~Turkington.
\newblock Ideal magnetofluid turbulence in two dimensions.
\newblock {\em J. Stat. Phys.}, 87:661, 1997.

\bibitem{AS}
M.~Ablowitz and H.~Segur.
\newblock On the evolution of packets of water waves.
\newblock {\em J. Fluid Mech.}, 92:691, 1979.

\bibitem{Pesceli}
H.~L. Pecseli.
\newblock Solitons and weakly nonlinear waves in plasmas.
\newblock {\em IEEE Trans. Plasma Sci}, 13:53, 1985.

\bibitem{HK}
A.~Hasegawa and Y.~Kodama.
\newblock {\em Solitons in Optical Communications}.
\newblock Oxford University Press, New York, 1995.

\bibitem{Hasimoto}
H.~Hasimoto.
\newblock  A soliton on a vortex filament.
\newblock {\em J. Fluid Mech.}, 139:67, 1972.

\bibitem{Majda}
A.~Majda.
\newblock Simplified asymptotic equations for slender vortex filaments.
\newblock {\em Proc. Symp. Appl. Math.}, 54:237, 1998.

\bibitem{LM}
Y.~Li and D.~McLaughlin.
\newblock Morse and Melnikov functions for NLS pdes.
\newblock {\em Commun. Math. Phys.}, 162:175, 1994.

\bibitem{ZS}
V.~E. Zakharov and A.~B. Shabat.
\newblock Exact theory of two--dimensional self--focusing and one--dimensional
  self--modulation of waves in nonlinear media.
\newblock {\em Soviet Phys. JETP}, 34:62, 1972.

\bibitem{AB}
D.~Anderson and M.~Bonnedal.
\newblock Variational approach to nonlinear self focusing of Gaussian laser
  beams.
\newblock {\em Phys. Fluids}, 22:105, 1979.

\bibitem{MNF}
D.~Mihalache, R.~G. Nazmitdinov, and V.~K. Fedyanin.
\newblock Nonlinear optical waves in layered structures.
\newblock {\em Soviet J. Part. Nucl.}, 20:86, 1989.

\bibitem{max}
C.~E. Max.
\newblock Strong self--focusing due to ponderomotive forces in plasmas.
\newblock {\em Phys. Fluids}, 19:74, 1976.

\bibitem{Zhidkov}
P.~E. Zhidkov.
\newblock On an invariant measure for a nonlinear Schr\"odinger equation.
\newblock {\em Soviet Math. Dokl.}, 43:431, 1991.

\bibitem{Bid}
B.~Bidegary.
\newblock Invariant measures for some partial differential equations.
\newblock {\em Physica D}, 82:340, 1995.

\bibitem{LRS}
J.~L. Lebowitz, H.~A. Rose, and E.~R. Speer.
\newblock Statistical mechanics of a nonlinear Schr\"odinger equation.
\newblock {\em J. Stat. Phys.}, 50:657, 1988.

\bibitem{Bourgain}
J.~Bourgain.
\newblock Periodic nonlinear Schr\"odinger equation and invariant measures.
\newblock {\em Commun. Math. Phys.}, 166:1, 1994.

\bibitem{strauss}
W.~A. Strauss.
\newblock {\em Nonlinear Wave Equations}.
\newblock AMS, Providence, RI, 1989.

\bibitem{McKean}
H.~P. McKean.
\newblock Statistical mechanics of nonlinear wave equations IV. Cubic
  Schr\"odinger.
\newblock {\em Commun. Math. Phys.}, 168:479, 1995.

\bibitem{PB}
M.~Plischke and B.~Bergersen.
\newblock {\em Equilibrium Statistical Physics}.
\newblock World Scientific, Singapore, 1994.

\bibitem{Jaynes}
E.~T. Jaynes.
\newblock Information theory and statistical mechanics.
\newblock {\em Phys.Rev.}, 106:620, 1957.

\bibitem{CHQZ}
C.~Canuto, M.~H. Hussaini, A.~Quateroni, and T.~A. Zang.
\newblock {\em Spectral Methods in Fluid Dynamics}.
\newblock Springer, Berlin-Heidelberg, 1988.

\bibitem{RevuzYor}
D.~Revuz and M.~Yor.
\newblock {\em Continuous Martingales and Brownian motion}.
\newblock Springer-Verlag, Berlin-Heidelberg, 1991.

\bibitem{Billingsley}
P.~Billingsley.
\newblock {\em Convergence of Probability Measures}.
\newblock Wiley, New York, 1968.

\bibitem{Adler}
R.~Adler.
\newblock {\em The Geometry of Random Fields}.
\newblock Wiley, New York, 1981.

\bibitem{CL}
T.~Cazenave and P.~L. Lions.
\newblock Orbital stability of standing waves for some nonlinear Schr\"odinger
  equations.
\newblock {\em Commun. Math. Phys.}, 85:549, 1982.

\bibitem{jtz2}
R.~Jordan, B.~Turkington, and C.~Zirbel.
\newblock On a mean-field statistical theory for a class of NLS equations.
\newblock {\em Los Alamos Technical Report}, 1999.

\bibitem{Adams}
R.~Adams.
\newblock {\em Sobolev Spaces}.
\newblock Academic Press, New York, 1975.

\end{thebibliography}

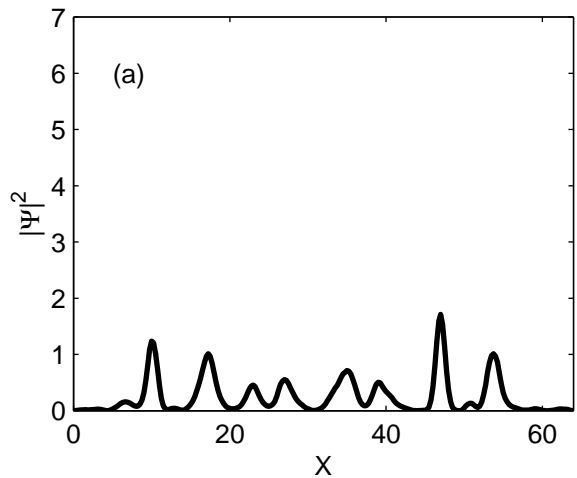
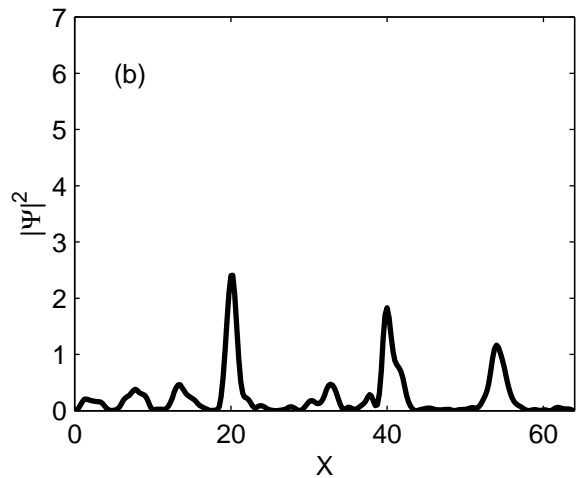
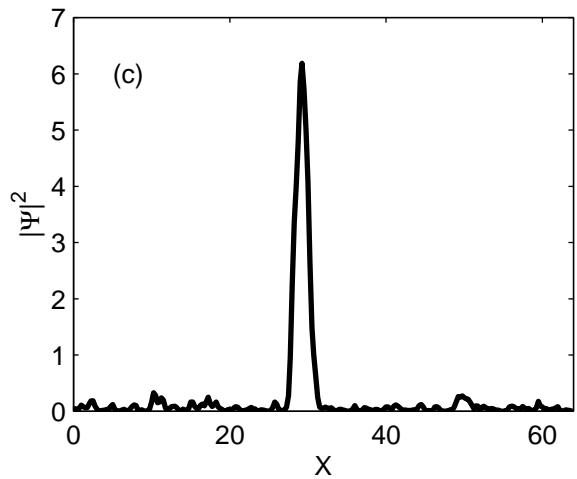
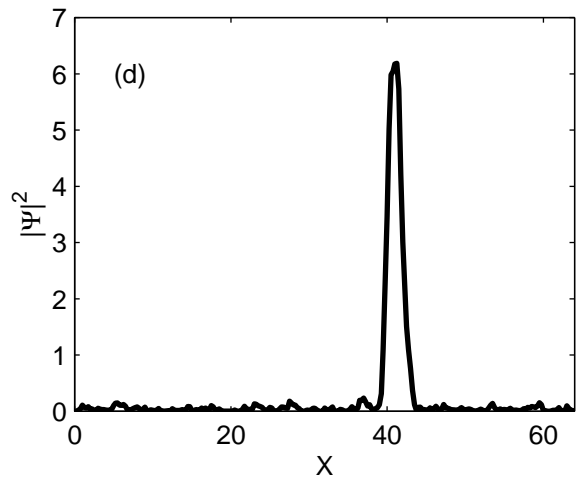
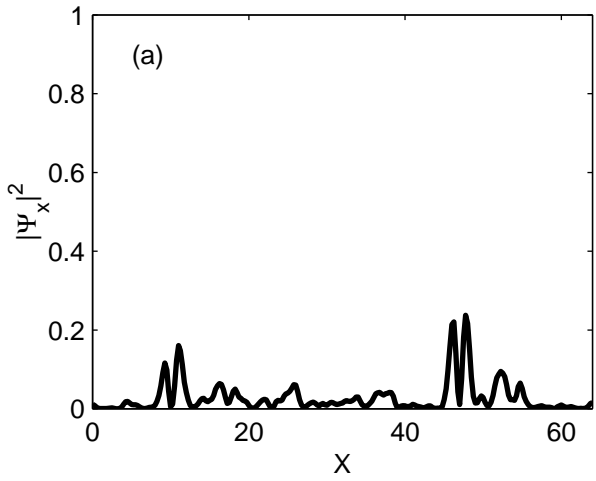
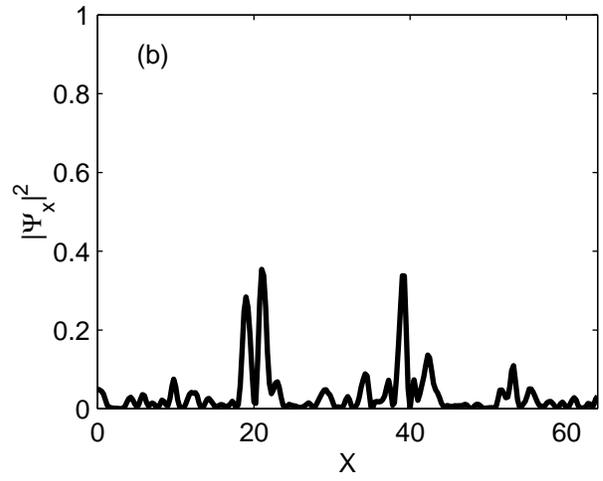
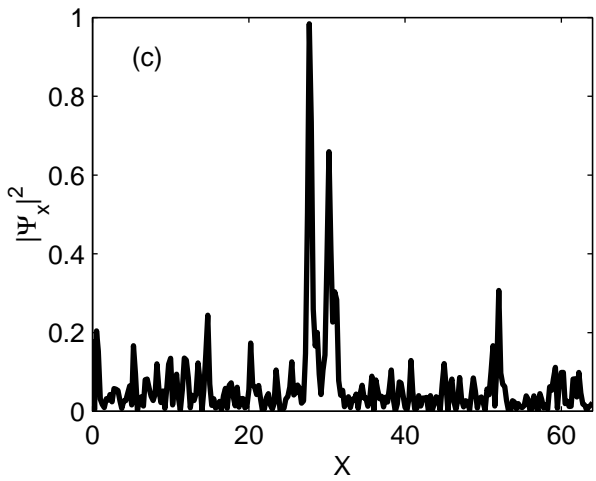
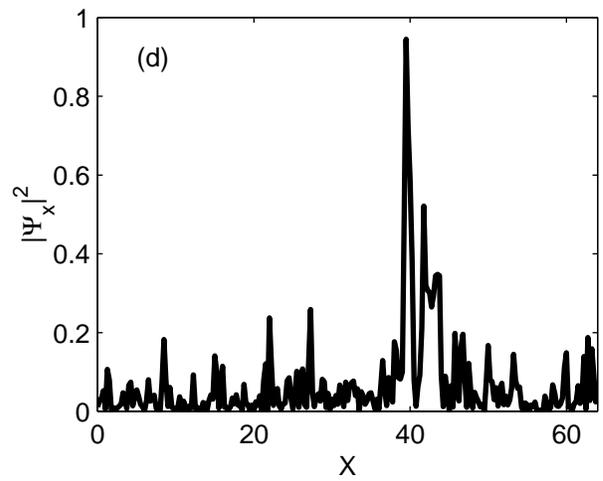
\end{document}